\documentclass[amsmath,prb,onecolumn]{revtex4}
\usepackage{amssymb,amsmath,amsthm,amsfonts,eucal,mathrsfs,amsbsy}
\usepackage{graphicx}
\usepackage{epstopdf}
\usepackage{color}
\begin{document}

%\title[The rise and fall of the amplitude through Exceptional Points]{The rise and fall of the amplitude through Exceptional Points: a Scattering matrix approach}

\title[EPs revealed by the integrated imaginary scattering eigenphase]{Exceptional Points revealed by the integrated imaginary scattering eigenphase}

\author{J. Col\'in-G\'alvez}
\email{colinlug05@gmail.com}
\affiliation{Departamento de F\'isica, Universidad Aut\'onoma Metropolitana-Iztapalapa,
	A. P. 55-534, 09340 Ciudad de M\'exico, Mexico.}

\author{J. E. Toral-Zavaleta}
\email{josetoralz00@gmail.com}
\affiliation{Departamento de F\'isica, Universidad Aut\'onoma Metropolitana-Iztapalapa,
	A. P. 55-534, 09340 Ciudad de M\'exico, Mexico.
}

\author{V. Dom\'inguez-Rocha\footnote{Corresponding author}}
\email{vdr@izt.uam.mx, vdr@azc.uam.mx, vidomr@gmail.com}
\affiliation{Departamento de F\'isica, Universidad Aut\'onoma Metropolitana-Iztapalapa,
	A. P. 55-534, 09340 Ciudad de M\'exico, Mexico, and\\
	Departamento de Ciencias B\'asicas, Universidad Aut\'onoma Metropolitana-Azcapotzalco,	Av. San Pablo 420, Col. Nueva Rosario, 02128, Ciudad de M\'exico, Mexico.
}

\vspace{10pt}
%\begin{indented}
%\item[]August 2017 (minor update March 2024)
%\end{indented}

\begin{abstract}
We propose and analytically demonstrate that the eigenphases of the scattering matrix provide direct, phase-sensitive signatures of exceptional points in open $\mathcal{PT}$-symmetric systems. Using a one-dimensional $\mathcal{PT}$-symmetric quantum dimer with balanced gain and loss, coupled to continuum leads, we track the evolution of the scattering eigenphases across the $\mathcal{PT}$-exact to $\mathcal{PT}$-broken transition. The imaginary parts of the eigenphases develop localized structures of opposite sign as the exceptional point is approached, reflecting the emergence of gain/loss asymmetry in the scattering eigenmodes. The exceptional point is located by the algebraic degeneracy condition of the $S$-matrix eigenvalues, and corroborated by two further, formally distinct analyses---the merging of the transmission resonances and the dynamics of the complex poles and zeros of $S(E)$---which place the transition within the same narrow parameter window, providing strong evidence for its location and physical reality. We then introduce the integrated imaginary scattering eigenphase $G(\gamma)$, which condenses the eigenphase information into a single experimentally accessible scalar and constitutes a fourth, resonance-free route to the same transition. $G(\gamma)$ is negligibly small in the $\mathcal{PT}$-exact phase and undergoes a sharp transition---a pronounced inflection followed by saturation into a plateau---near the exceptional point. The inflection point of $G(\gamma)$ precedes $\gamma_{\textrm{EP}}$ and coincides with the maximum amplitude of the transmission resonances, revealing that peak gain/loss asymmetry and eigenstate coalescence are governed by independent conditions in open systems---a direct fingerprint of their finite coupling to the continuum. Because the analysis is formulated entirely at the level of the $S$-matrix---without reference to any specific physical realization---the results are universally applicable across wave systems, including photonic, acoustic, and microwave platforms where phase-resolved measurements are available.
\end{abstract}

\maketitle

\vspace{2pc}

\section{Introduction}

Parity–time ($\mathcal{PT}$)-symmetric systems occupy a distinctive place in non-Hermitian physics, sustaining entirely real spectra despite being non-Hermitian---up to a critical value of the non-Hermitian parameter where eigenstates coalesce at an exceptional point (EP)~\cite{Bender1998,Bender2002,Bender2007,Kato}. EPs appear when the variation of that parameter allows the coalescence of two (or more) naturally repelling eigenvalues, as well as their eigenvectors, of the associated Hamiltonian operator~\cite{Alu2021, Heiss}. This transition between exact and broken $\mathcal{PT}$-symmetric phases has been extensively studied across electromagnetism~\cite{electromagnetism2020}, optics~\cite{optics2017, Chong2011, Ramezani2011}, microwaves~\cite{Dietz}, acoustics~\cite{acoustics2016}, photonics~\cite{photonics2019}, elasticity~\cite{vdomEP}, quantum mechanics~\cite{Schomerus, quantummechanics2018, Ortega, Dorey, Ruschhaupt, Noble}, electronics~\cite{electronics2017, Kottos2011} and atomic physics~\cite{atomicphysics2007}. EPs have been applied in improving the sensitivity of optical measurements~\cite{optics2017}, in robust transfer of energy wireless~\cite{electronics2017}, in unidirectional invisibility~\cite{atomicphysics2007, Chong2010}, slowing light~\cite{Moiseyev2018}, in alteration of spectral bands~\cite{Zhen2015, Kartashov2018}, or in accelerometers~\cite{KottosNat}.

In open systems, however, exceptional points are most directly probed through scattering experiments. The scattering matrix provides a natural description for such physical systems, encoding all measurable transport and resonant properties~\cite{vdomEP,Schomerus,quantummechanics2018}. Accordingly, $\mathcal{PT}$-symmetric scattering systems have been widely investigated using amplitude-based observables---such as transmission peaks, reflection zeros, resonance linewidths, or pole trajectories in the complex plane---often within tight-binding models or periodic potentials, as indicators of non-Hermitian behavior and EP signatures~\cite{optics2017, Ramezani2011, Ortega, Hatano2015, Hatano2021, Izrailev2015, Bermudez2020, Valle, Znojil, Tielas, Vemuria, Longhi2011}. These approaches are well established, and we employ several of them here to locate the EP of our system with high precision. Their practical limitation is that they require resolving and tracking individual spectral features---distinct transmission peaks, or individual poles continued to complex energy---a task that becomes increasingly demanding when resonances broaden, overlap, or merge into the continuum. In open non-Hermitian systems, this motivates the search for a complementary characterization of the $\mathcal{PT}$ transition that does not rely on identifying individual resonant states.

By contrast, the role of scattering phases has remained comparatively underexplored, despite being directly accessible in modern phase-resolved experiments such as those employing vector network analyzers in photonics~\cite{Luna}, microwave~\cite{PAbsorption}, or even elastodynamics~\cite{vdomEP, vdom2020} systems. A systematic, phase-based signature of EPs rooted directly in the scattering matrix itself is therefore lacking. Here, we propose and analytically demonstrate that the eigenphases of the scattering matrix provide a direct and robust phase-sensitive signature of exceptional points. The eigenvalues of the scattering matrix are complex numbers of the form $e^{\textrm{i}\theta}$, where $\theta=\theta^{\textrm{Re}} +\textrm{i}\theta^{\textrm{Im}}$ is a complex phase whose real part encodes the usual phase shift and whose imaginary part governs global amplification or attenuation, depending on its sign. By tracking the evolution of these eigenphases across the $\mathcal{PT}$-symmetry-breaking transition, we identify a clear fingerprint of the EP in the imaginary parts $\theta_n^{\textrm{Im}}$, which develop localized structures of opposite sign as the exceptional point is approached. We further introduce the integrated imaginary scattering eigenphase $G(\gamma)$, a single experimentally accessible scalar that captures the $\mathcal{PT}$ transition globally---without requiring energy-resolved measurements or the identification of individual resonance peaks. These signatures are intrinsic to the scattering description and do not rely on Hamiltonian eigenmodes or bound-state spectra.

To illustrate these ideas concretely, we analyze a minimal yet physically relevant model: a one-dimensional $\mathcal{PT}$-symmetric open quantum dimer coupled to continuum leads. The system consists of two identical scatterers with balanced gain and loss and admits an exact analytical formulation of the full scattering matrix. This allows us to trace the scattering eigenphases as functions of both energy and the non-Hermitian strength $\gamma$. Crucially, because our analysis is formulated entirely at the level of the scattering matrix---without recourse to tight-binding approximations or periodic potentials---the resulting eigenphase signatures are not restricted to this specific geometry, but follow directly from the scattering-matrix description of open systems. They are therefore applicable to any wave system describable by an $S$-matrix, including quantum, photonic, and microwave platforms~\cite{Luna,Mello1993, Saphiro, Moises2017, Prosen, Caio, MelloBook}. In this way, scattering eigenphases emerge as a universal and experimentally accessible diagnostic of exceptional points in open non-Hermitian systems.

The rest of the paper is organized as follows. Section~\ref{sec:Sec2} introduces the $\mathcal{PT}$-symmetric dimer and derives its exact scattering matrix and eigenphases. Section~\ref{sec:EPlocation} locates the exceptional point through the algebraic degeneracy of the $S$-matrix eigenvalues, and corroborates it with the evolution of the scattering resonances and the dynamics of the complex poles and zeros of $S(E)$. Section~\ref{sec:eigenphases-Ggamma} analyzes the scattering eigenphases across the $\mathcal{PT}$-symmetry-breaking transition and introduces the integrated imaginary eigenphase $G(\gamma)$ as a resonance-free diagnostic consistent with the exceptional-point location established in Sec.~\ref{sec:EPlocation}. Section~\ref{sec:conclusions} summarizes the main results and discusses their implications for experimental platforms. The analytical expressions for the scattering matrix elements are collected in the Appendices for completeness and reproducibility.

%%%%%%%%%%%%%%%%%%%%%%%%%%%%%%%%%%%%%%%%%%%%%%%%%%%%%%%%%%%%%%%%%%%%%%%%%%%%%%%%%%%%%%%%%%%%%%%%%%%%%%%%%%%%%%%%%%%%%%%%%%%%%%%%%%%%%%%%%%%%%%%%%%%%%%%%%%%%%%%%%%%%%%%%%%%%%%%%%%%%%%%%%%%%%%%%%%%%%%%%%%%%%%%%

\section{The $S$ matrix of a $\mathcal{PT}$-symmetric dimer}
\label{sec:Sec2}

\subsection{$\mathcal{PT}$-symmetric open quantum dimer}
\label{subsec:model}

\begin{figure}
	\centering
	\includegraphics[width=11cm]{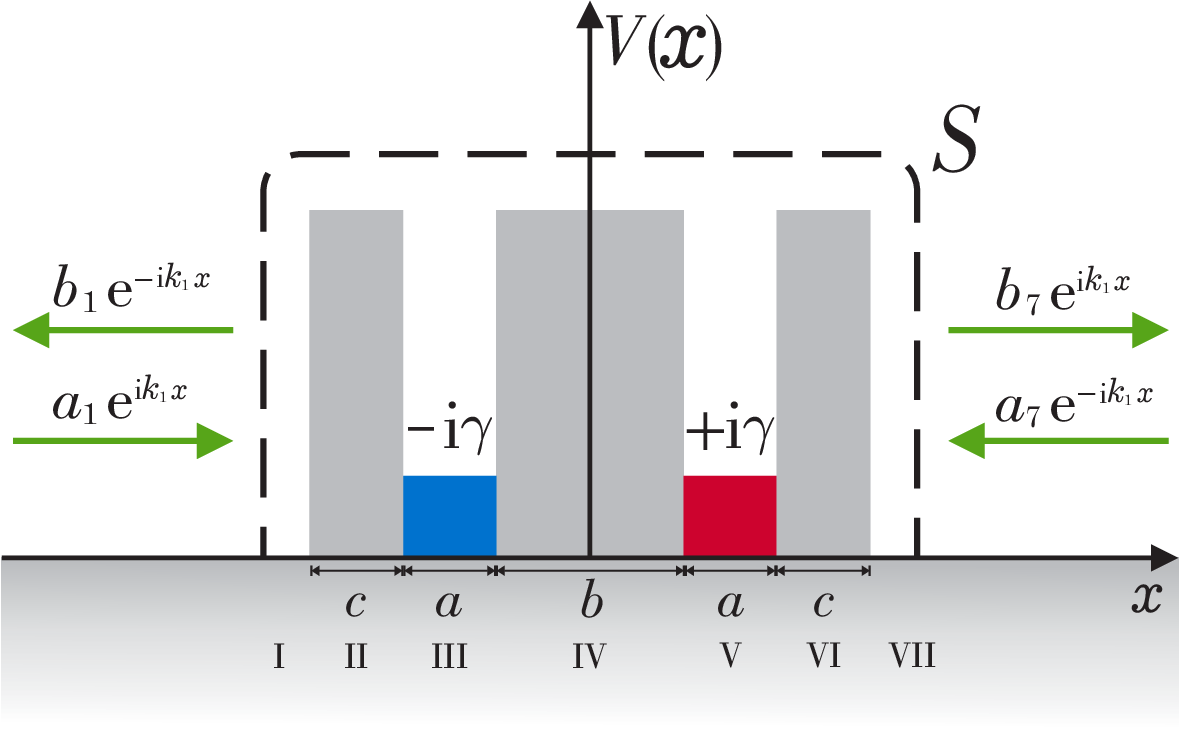}
	\caption{(Color online) Diagram of the studied dimer composed of two scattering units, each composed of three potential barriers. The potential of the outer barriers of each scattering unit is real, while the potential described by the central barrier is complex. The sign of the imaginary part of the central barrier gives rise to loss (blue region) or gain (red region). This system is invariant under the combined action of parity and time reversal, and is therefore $\mathcal{PT}$-symmetric.}
	\label{fig1:Diagram}
\end{figure}

We consider a one-dimensional quantum dimer with balanced gain and loss, coupled to continuum leads on both sides, as illustrated in Fig.~\ref{fig1:Diagram}. The dimer consists of two identical scattering units arranged contiguously along the propagation direction, each one with a real–complex–real structure. When placed back-to-back, this arrangement produces a central real-potential region of twice the width of the outer barriers. The real parts of the potentials are identical for both units, while the imaginary parts have equal magnitude but opposite signs, introducing spatially separated loss and gain regions. The non-Hermitian strength is controlled by a single parameter $\gamma$, which governs the magnitude of gain and loss and serves as the control parameter for the $\mathcal{PT}$ transition.

%%%%%%%%%%%%%%%%%%%%%%%%
The non-Hermitian regions are characterized by complex wave numbers, $k_3=K+\textrm{i}\Gamma$ in the loss region and $k_5=K-\textrm{i}\Gamma$ in the gain region, with
\begin{equation}
	K=\sqrt{\frac{m}{\hbar^2}\left[(E-V')+\sqrt{(E-V')^2+\gamma^2}\right]},
	\qquad
	\Gamma=\sqrt{\frac{m}{\hbar^2}\left[\sqrt{(E-V')^2+\gamma^2}-(E-V')\right]},
	\label{eq:K-Gamma}
\end{equation}
where both $K$ and $\Gamma$ are real and positive. The oscillatory part of the wave is governed by $K$, while $\Gamma$ controls the exponential attenuation in the loss region and the corresponding amplification in the gain region, and therefore sets the strength of the non-Hermitian effect\footnote{Not to be confused with the resonance linewidth $\Gamma_{w,n}$ introduced in Sec.~\ref{subsec:poles-zeros}.}. The remaining wave numbers, together with the stationary solutions in the seven regions, are given in Appendix~A.
%%%%%%%%%%%%%%%%%%%%%%%

The system is invariant under the combined action of parity (spatial reflection about the center of the dimer) and time reversal, and is therefore $\mathcal{PT}$-symmetric by construction. For $\gamma = 0$ the system is Hermitian\footnote{In this limit $r=r'$ and $t=t'$, as a consequence of specular reflection and time-reversal symmetry.} and supports well-defined scattering resonances. As $\gamma$ increases, pairs of resonant states approach each another and coalesce at an exceptional point, beyond which the system enters the $\mathcal{PT}$-broken phase.

Particles incident from the left or right are described by plane waves, and transport is fully characterized by the scattering matrix ($S$ matrix), which relates outgoing ($b_1$ and $b_7$) to incoming ($a_1$ and $a_7$) wave amplitudes in the two asymptotic channels~\cite{MelloBook}. For a one-dimensional system with two leads, $S$ is a $2\times2$ matrix,
\begin{equation}
	S=\begin{bmatrix}
		r & t'\\
		t & r'    
	\end{bmatrix},
	\label{ec2:Smatrix}
\end{equation}
where $r$ ($r'$) and $t$ ($t'$) denote reflection and transmission amplitudes for incidence from the left (right), respectively. The dependence on $\gamma$ enters through the complex wave numbers in the gain and loss regions given in Eq.~\ref{eq:K-Gamma}.

%%%%%%%%%%%%%%%%%%%%%%%%%%%%%%%%%%%%%%%%%%%%%%%%%%%%%%%%%%%%%%%%%%%%%%%%%%%%%%%%%%%%%%%%%%%%%%%%%%%%%%%%
%%%%%%%%%%%%%%%%%%%%%%%%%%%%%%%%%%%%%%%%%%%%%%%%%%%%%%%%%%%%%%%%%%%%%%%%%%%%%%%%%%%%%%%%%%%%%%%%%%%%%%%%

\subsection{Exact scattering matrix and eigenphases}

The scattering properties of the $\mathcal{PT}$-symmetric dimer are fully encoded in its $S$ matrix, which relates the amplitudes of outgoing and incoming waves in the asymptotic leads as in Eq.~(\ref{ec2:Smatrix}). Its elements are obtained by imposing continuity of the wave function and its derivative at each interface between regions, leading to an equations system that connects the outgoing and incoming amplitudes. Solving this matching problem yields exact analytical expressions for the reflection and transmission amplitudes $r$, $r'$, $t$, and $t'$ as functions of the incident energy $E$ and the non-Hermitian parameter $\gamma$. While these expressions are lengthy, they can be derived in closed form and are reported in Appendix~A for completeness.

In $\mathcal{PT}$-symmetric open systems, the scattering matrix satisfies a generalized symmetry relation that constrains the eigenvalues of $S$. In contrast to Hermitian systems, where the eigenvalues of $S$ lie on the unit circle, non-Hermitian gain and loss allow them to move off the unit circle while still obeying symmetry-imposed pairing relations. We therefore focus on the eigenvalues $\lambda_{1,2}$ of the scattering matrix, which can be written in the form
\begin{equation}
	\lambda_n = e^{\textrm{i}\theta_n},
	\label{ec:lambda_n}
\end{equation}
where $\theta_n$ are, in general, complex quantities. The real part $\textrm{Re}\,\theta_n$ represents the scattering phase shift, while the imaginary part $\textrm{Im}\,\theta_n$ accounts for net amplification or attenuation associated with gain and loss.

The evolution of these eigenphases as functions of energy and non-Hermiticity constitutes the central object of this work. In particular, their behavior near the $\mathcal{PT}$-symmetry-breaking transition reveals clear signatures of the exceptional point. Before turning to the eigenphases in Sec.~\ref{sec:eigenphases-Ggamma}, we first locate the exceptional point of this system in the next section.

%%%%%%%%%%%%%%%%%%%%%%%%%%%%%%%%%%%%%%%%%%%%%%%%%%%%%%%%%%%%%%%%%%%%%%%%%%%%%%%%%%%%%%%%%%%%%%%%%%%%%%%%%%%%%%%%%%%%%%%%%%%%%%%%%%%%%%%%%%%%%%%%%%%%%%%%%%%%%%%%%%%%%%%%%%%%%%%%%%%%%%%%%%%%%%%%%%%%%%%%%%%%%%%%%%%%%%%%%%%%%%%%%%%%%%%%%%%%%%%%%%%%%%%%%%%%%%%%%%%%%%%%%%%%%%%%%%%%%%%%%%%%%%%%%%%%%%%%%%%%%%%%%%%%%%%%
\section{Scattering resonances and identification of the exceptional point}
\label{sec:EPlocation}
%%%%%%%%%%%%%%%%%%%%%%%%%%%%%%%%%%%%%%%%%%%%%%%%%%%%%%%%%%%%%%%%%%%%%%%%%%%%%%%%%%%%%%%%%%%%%%%%%%%%%%%%%%%%%%%%%%%%%%%%%%%%%%%%%%%%%%%%%%%%%%%%%%%%%%%%%%%%%%%%%%%%%%%%%%%%%%%%%%%%%%%%%%%%%%%%%%%%%%%%%%%%%%%%

\subsection{Resonance-peak tracking}
\label{subsec:EP_identification}

\begin{figure}
	\centering
	\includegraphics[width=10cm]{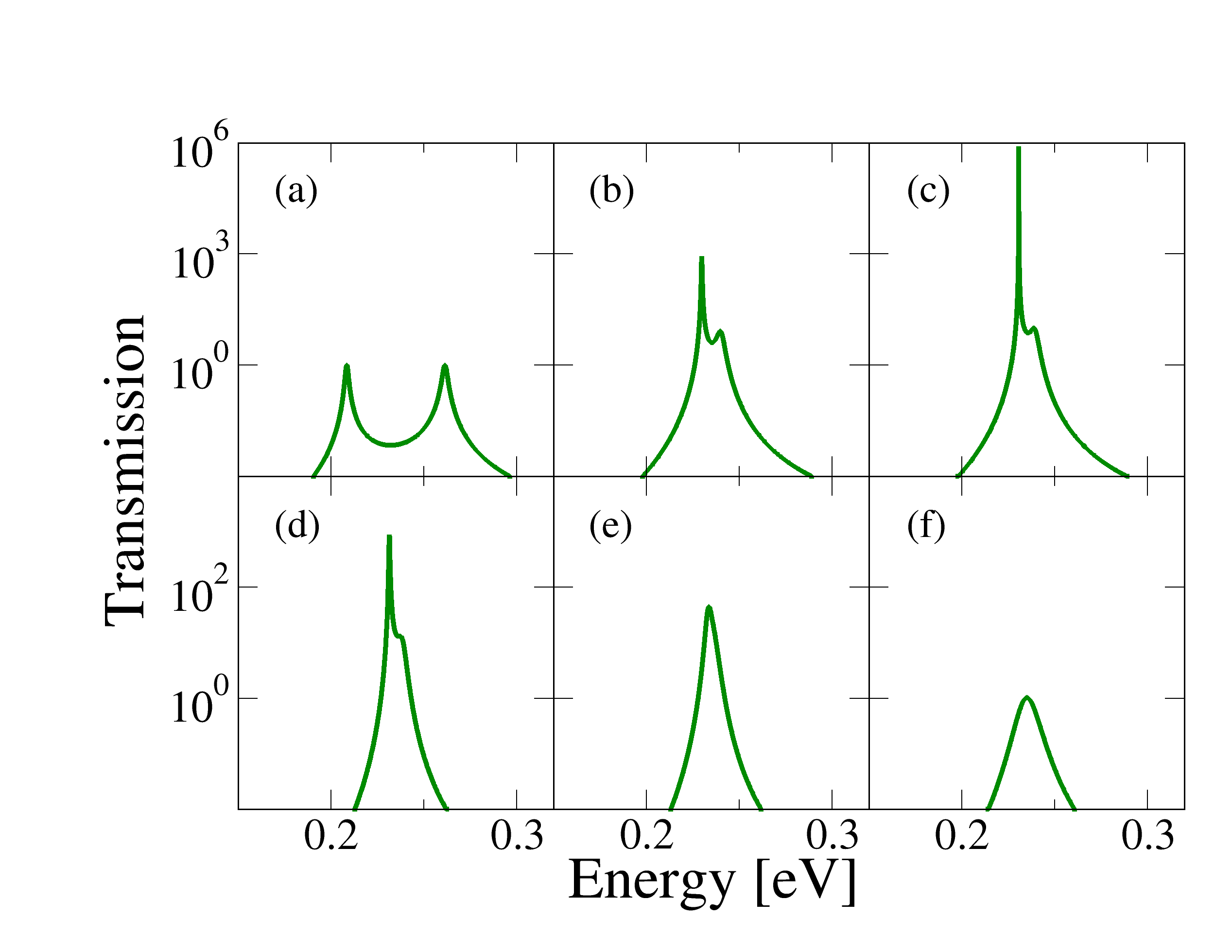}
	\caption{(Color online) Log-scale transmission spectrum as a function of energy in which only the first doublet of the dimer is plotted. Dimer parameters, for this and subsequent figures, are $a=1.15~$nm, $b=0.01~\text{nm}$, $c=0.02~\text{nm}$, $V_\textrm{b}= 50~\text{eV}$, and $V'=0~\text{eV}$. (a) For the Hermitian case ($\gamma=0$) the maximum values of the resonances are found at $E_{1,1}=0.2086~\text{eV}$ (the first subscript refers to the associated doublet while the second one refers to the associated resonance), and $E_{1,2}=0.2615~\text{eV}$. For $\gamma=0$ the resonances have a unit amplitude and a maximum separation. As we increase the value of $\gamma$ the separation between the resonances decreases, as observed for $\gamma$ equal to (b) 0.0264, (c) 0.026564, and (d) 0.0267, all below $\gamma_{\textrm{EP}}$. The resonance amplitude does not grow monotonically in this range. It reaches an extremely sharp, near-singular maximum at $\gamma=0.026564$, and has already decreased substantially by panel (d). Beyond the exceptional point ($\gamma_{\textrm{EP}}\approx0.02685$), only a single, fully merged resonance persists, with its width increasing and its amplitude continuing to decrease, as observed in panels (e) for $\gamma=0.0270$ and (f) for $\gamma=0.0280$. The same values of $\gamma$ are used in figures~\ref{fig4:ReThetas} and \ref{fig5:ImThetas}.}
	\label{fig2:TransmissionSpectrum}
\end{figure}

The $\mathcal{PT}$-symmetric dimer introduced in Sec.~\ref{subsec:model} supports scattering resonances that manifest as sharp peaks in the transmission spectrum $T=|t|^2$. These resonances encode the spectral properties of the open system and evolve continuously as the gain–loss strength $\gamma$ is varied. As shown in Fig.~\ref{fig2:TransmissionSpectrum}, for small $\gamma$ the transmission exhibits well-separated resonant peaks corresponding to quasi-bound states of the dimer.

\begin{figure}
	\centering
	\includegraphics[width=10cm]{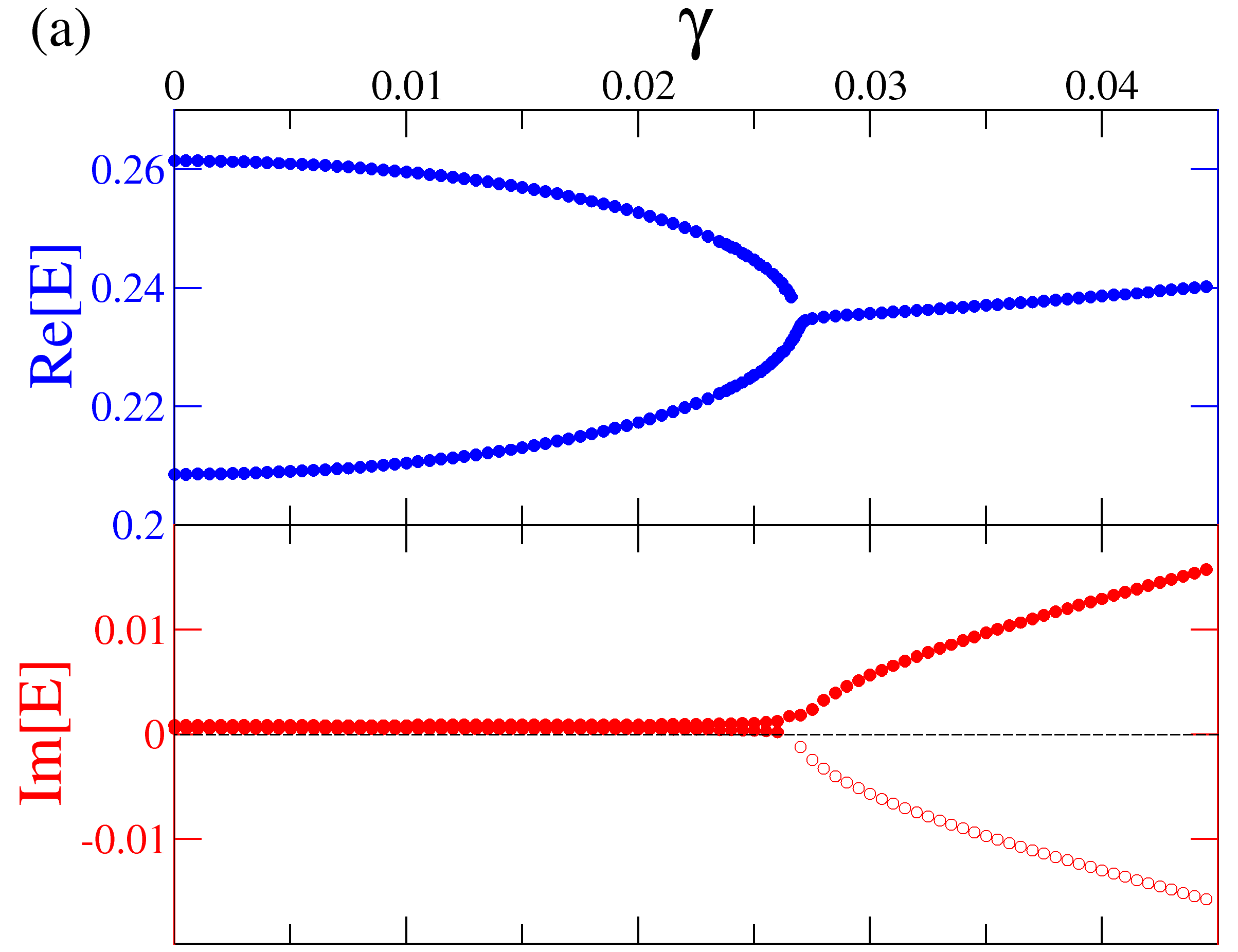}
	\vspace{6pt}
	\includegraphics[width=10cm]{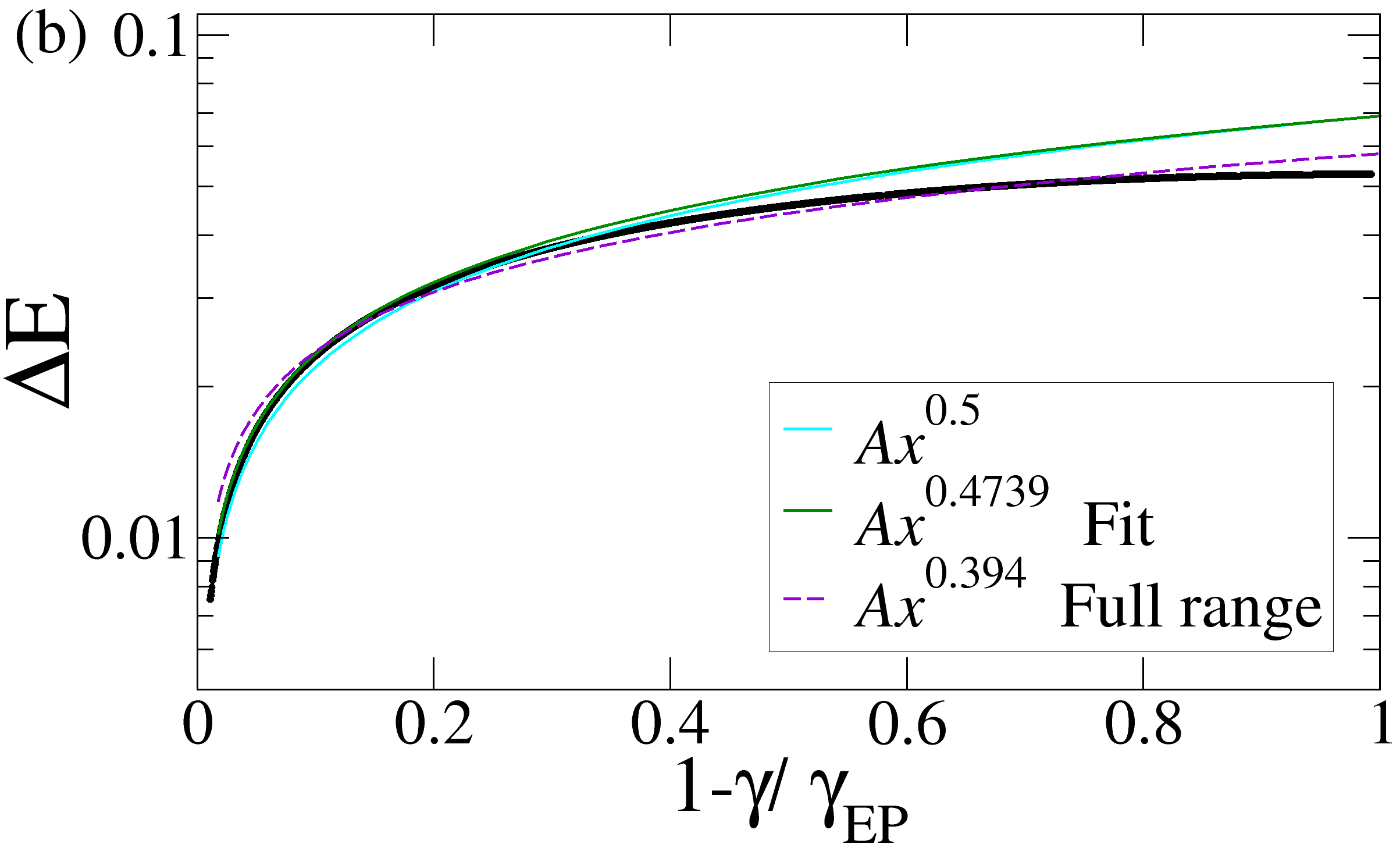}
	\caption{(Color online) (a) Evolution of the scattering resonances of the $\mathcal{PT}$-symmetric dimer as a function of $\gamma$. The real part (upper panel) tracks the resonance positions. For $\gamma=0$ the doublet is maximally separated due to level repulsion, and the separation decreases monotonically until both resonance peaks merge at the exceptional point. Just below $\gamma_{\textrm{EP}}$ only the lower-energy resonance is numerically resolvable, which accounts for the gap in the higher-energy trajectory. Beyond the EP a single merged resonance persists. The imaginary part (lower panel) tracks the resonance widths, obtained as half the full width at half maximum of $T(E)$ (see Sec.~\ref{subsec:EP_identification}). As $\gamma$ increases the resonances broaden and eventually coalesce into a single significantly wider feature. Open circles in the negative region of $\textrm{Im}[E]$ are reflections of the positive values~\cite{Kottos2011}. The dashed line marks $\textrm{Im}[E]=0$. (b) Energy splitting $\Delta E$ as a function of $1-\gamma/\gamma_{\textrm{EP}}$ (black), together with power-law fits $\Delta E=A(1-\gamma/\gamma_{\textrm{EP}})^{\beta}$: a fit restricted to $1-\gamma/\gamma_{\textrm{EP}}\le0.1$ (dark green), yielding $A=0.069$ and $\beta=0.474$ ($R^2=0.999997$), closely matching the theoretical square-root exponent $\beta=0.5$ (cyan); and, for comparison, a fit over the full range shown (violet, dashed), yielding a smaller effective exponent $\beta=0.394$ ($R^2=0.983$). The difference illustrates that the asymptotic square-root behavior is recovered only sufficiently close to $\gamma_{\textrm{EP}}$.}
	\label{fig:EPtrajectory}
\end{figure}

To track the evolution of these resonances, we extract their positions from the maxima of $T$ as a function of $\gamma$. These maxima are obtained by direct numerical maximization of the exact transmission spectrum, without recourse to any line-shape fitting. The corresponding widths are obtained from the same exact curve, by locating the two energies at which $T$ falls to half of its peak value and identifying $\textrm{Im}[E]$ with half of the resulting full width at half maximum, $\textrm{Im}[E]=\Gamma_{w}/2$. This is the standard Lorentzian relation, applied here as a direct geometric measurement on the numerically exact transmission rather than through an explicit Lorentzian fit. For $\gamma=0$, the system is Hermitian and the resonances form doublets associated with symmetric (lower-energy resonance) and antisymmetric (higher-energy resonance) scattering states. As $\gamma$ increases, each pair approaches one another in energy, reflecting the progressive non-Hermitian coupling induced by balanced gain and loss. This evolution is summarized in Fig.~\ref{fig:EPtrajectory}(a), where the trajectories of the resonant energies ($\textrm{Re}[E]$), as well as their widths ($\textrm{Im}[E]$), are plotted against $\gamma$. At a critical value $\gamma=\gamma_{\textrm{EP}}\approx0.02685$, two resonances coalesce into a single scattering singularity, signaling the presence of an exceptional point. The peaks of $T$ merge at this value of $\gamma$, while the underlying poles approach without strictly coalescing (see Sec.~\ref{subsec:poles-zeros}). Beyond this point, the merged resonance evolves into a broadened spectral feature, marking the onset of the $\mathcal{PT}$-broken phase. The distinctive square-root characteristic shape of an exceptional point is confirmed in Fig.~\ref{fig:EPtrajectory}(b), which shows the energy splitting $\Delta E$ as a function of $1-\gamma/\gamma_{\textrm{EP}}$. The data follow a power-law behavior $\Delta E = A(1-\gamma/\gamma_{\textrm{EP}})^{\beta}$; a fit restricted to a narrow window near the exceptional point ($1-\gamma/\gamma_{\textrm{EP}}\le0.1$) yields $A=0.069\pm0.0001$ and $\beta=0.474\pm0.0001$ ($R^2=0.999997$), close to the theoretical value $1/2$, while a fit over the full accessible range yields a smaller effective exponent, $\beta=0.394\pm0.003$ ($R^2=0.983$), reflecting the fact that the asymptotic square-root law holds only sufficiently close to $\gamma_{\textrm{EP}}$. The residual deviation from $\beta=1/2$ in the near-EP window reflects the finite natural linewidths of the resonances (Sec.~\ref{subsec:poles-zeros}).

The identification of the exceptional point through the evolution of scattering resonances provides a clear and physically transparent picture of the $\mathcal{PT}$-symmetry-breaking transition in the open dimer. However, this analysis relies on the ability to resolve and track individual transmission peaks as system parameters are varied. In open non-Hermitian systems, resonances may broaden, overlap, or merge into the continuum, making their extraction increasingly ambiguous and system-dependent. This motivates the search for characterizations of the $\mathcal{PT}$ transition that do not rely on the identification of individual resonant states, which we develop in the remainder of this section.

\subsection{Algebraic confirmation via $S$-matrix eigenvalue degeneracy}
\label{subsec:algebraic-EP}

Within the $S$-matrix eigenvalue framework~\cite{Ambichl} (see Eq.~(\ref{ec2:matriz_diagonalizada})), an exceptional point occurs where the two eigenvalues of $S$ become degenerate at a physically incident real energy, i.e., where the discriminant appearing in Eq.~\eqref{ec2:matriz_diagonalizada} vanishes,
\begin{equation}
	(r-r')^2+4tt' = 0,
	\label{eq:disc-again}
\end{equation}
with $r$, $r'$, $t$, $t'$ evaluated at real $E$ and real $\gamma$. Solving Eq.~\eqref{eq:disc-again} numerically, seeded in the immediate neighborhood of $\gamma_{\textrm{EP}}=0.02685$ obtained from resonance-peak tracking (Sec.~\ref{subsec:EP_identification}), yields
\begin{equation}
	P_{\textrm{EP}} = (E_{\textrm{EP}},\gamma_{\textrm{EP}}), \qquad	E_{\textrm{EP}} = 0.2325370594~\textrm{eV}, \qquad	\gamma_{\textrm{EP}} = 0.0268500004,
	\label{eq:EP-algebraic-value}
\end{equation}
with the two eigenvalues degenerate to within $|\lambda_1-\lambda_2|\sim10^{-14}$, and $|t|\approx12.3\neq0$ at this point, confirming that this is a genuine exceptional point rather than a trivial degeneracy (for which $t=0$ identically; see the discussion below). This value reproduces the peak-tracking estimate $\gamma_{\textrm{EP}}\approx0.02685$ used to seed the search, and refines it by six further significant figures (shown directly in Fig.~\ref{fig:poles-real-part}). We verified Eq.~\eqref{eq:EP-algebraic-value} independently using two separate numerical implementations of the analytical $S$-matrix (Appendix~A), obtaining agreement to at least ten significant figures.

A technical remark is in order regarding the nature of this degeneracy condition. For a general $2\times2$ matrix, the vanishing of the discriminant in Eq.~\eqref{eq:disc-again} guarantees equal eigenvalues, but the matrix is non-diagonalizable---i.e., a genuine exceptional point, with coalescing eigenvectors---only if the off-diagonal elements $t$, $t'$ do not vanish simultaneously at that point; if they do, the discriminant vanishes trivially and the matrix reduces to a multiple of the identity, with no associated eigenvector coalescence. In the Hermitian limit ($\gamma=0$), the dimer symmetry enforces $r=r'$ and $t=t'$ (Sec.~\ref{subsec:model}), so that Eq.~\eqref{eq:disc-again} reduces identically to $4t^2=0$. Any zero of the discriminant at $\gamma=0$ is therefore automatically a transmission zero, $t=0$, and corresponds to the trivial case rather than to a genuine coalescence of resonant states. This symmetry is broken once $\gamma\neq0$, so that Eq.~\eqref{eq:disc-again} and $t=0$ become independent conditions, and the non-trivial root reported in Eq.~\eqref{eq:EP-algebraic-value}, with $t\neq0$, signals a genuine exceptional point.

\subsection{Complex pole and zero dynamics}
\label{subsec:poles-zeros}

A complementary, and formally distinct, characterization of non-Hermitian degeneracy is available from the analytic structure of $S(E)$ continued to complex energy, where resonances appear as poles of $t(E)$, $E_{\textrm{pole},n}=E_{R,n}-\textrm{i}\Gamma_{w,n}/2$, and an exceptional point in the sense of Kato~\cite{Kato} corresponds to the coalescence of two such poles into a single, non-diagonalizable double pole~\cite{Hernandez2011}. We examine this structure for the two poles $p_1$, $p_2$ of the first doublet, tracked by continuation in $\gamma$ from the exactly known Hermitian-limit values ($E_{1,1}=0.2086$~eV, $E_{1,2}=0.2615$~eV), together with the corresponding zeros $z_1$, $z_2$ of $\det S(E)$. As shown in Sec.~\ref{subsec:Subsec31}, the reciprocal pairing $\lambda_1\lambda_2^*=1$ forces each zero to sit, at every $\gamma$, at the exact Schwarz-conjugate position of its partner pole, $z_n=p_n^*$, since $\det S=\lambda_1\lambda_2$ and a diverging eigenvalue forces the companion eigenvalue toward zero at the same point.

\begin{figure}[htbp]
	\centering
	\includegraphics[width=10cm]{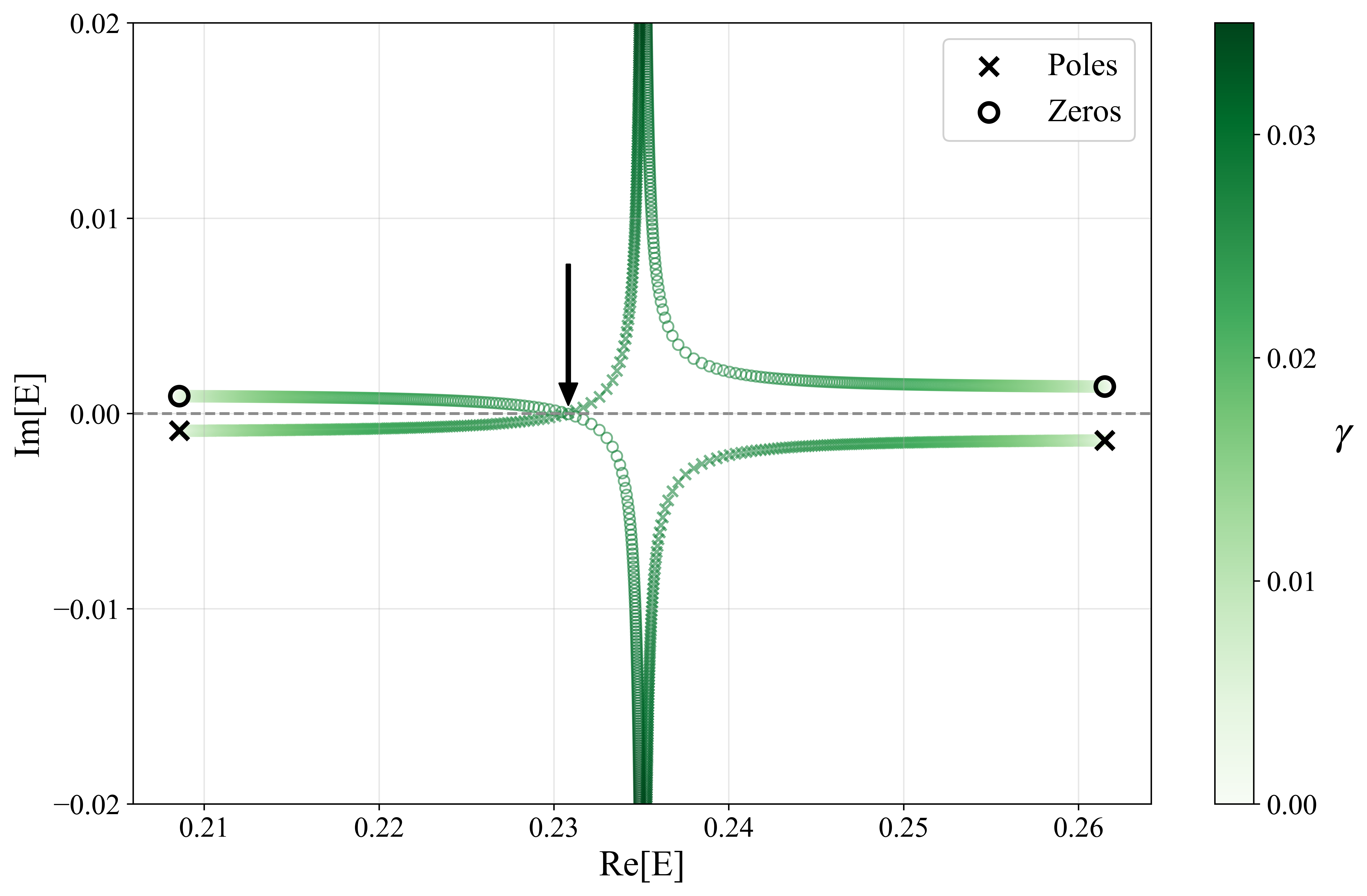}
	\caption{Trajectories of the two resonance poles ($p_1$, $p_2$) and their Schwarz-conjugate zeros ($z_1$, $z_2$) of $\det S(E)$ in the complex energy plane as $\gamma$ increases from 0 to 0.035. Only the lower-energy pair, $p_1$ and $z_1$, crosses the real axis, at $\gamma_{max}=0.026583$. The higher-energy pair, $p_2$ and $z_2$, instead diverges monotonically away from the real axis.}
	\label{fig:poles-complex-plane}
\end{figure}

Figure~\ref{fig:poles-complex-plane} shows the resulting trajectories. Before the transition, $p_1$ and $z_1$ (respectively $p_2$ and $z_2$) coincide, in real part, with the two resonance peaks of Sec.~\ref{subsec:EP_identification}, independently cross-checking the pole-tracking procedure. Two qualitatively distinct behaviors are visible as $\gamma$ increases.

%--------------------------------------------------------------
% FIGURA B (Eduardo): Re(E) vs gamma
%--------------------------------------------------------------
\begin{figure}[htbp]
	\centering
	\includegraphics[width=10cm]{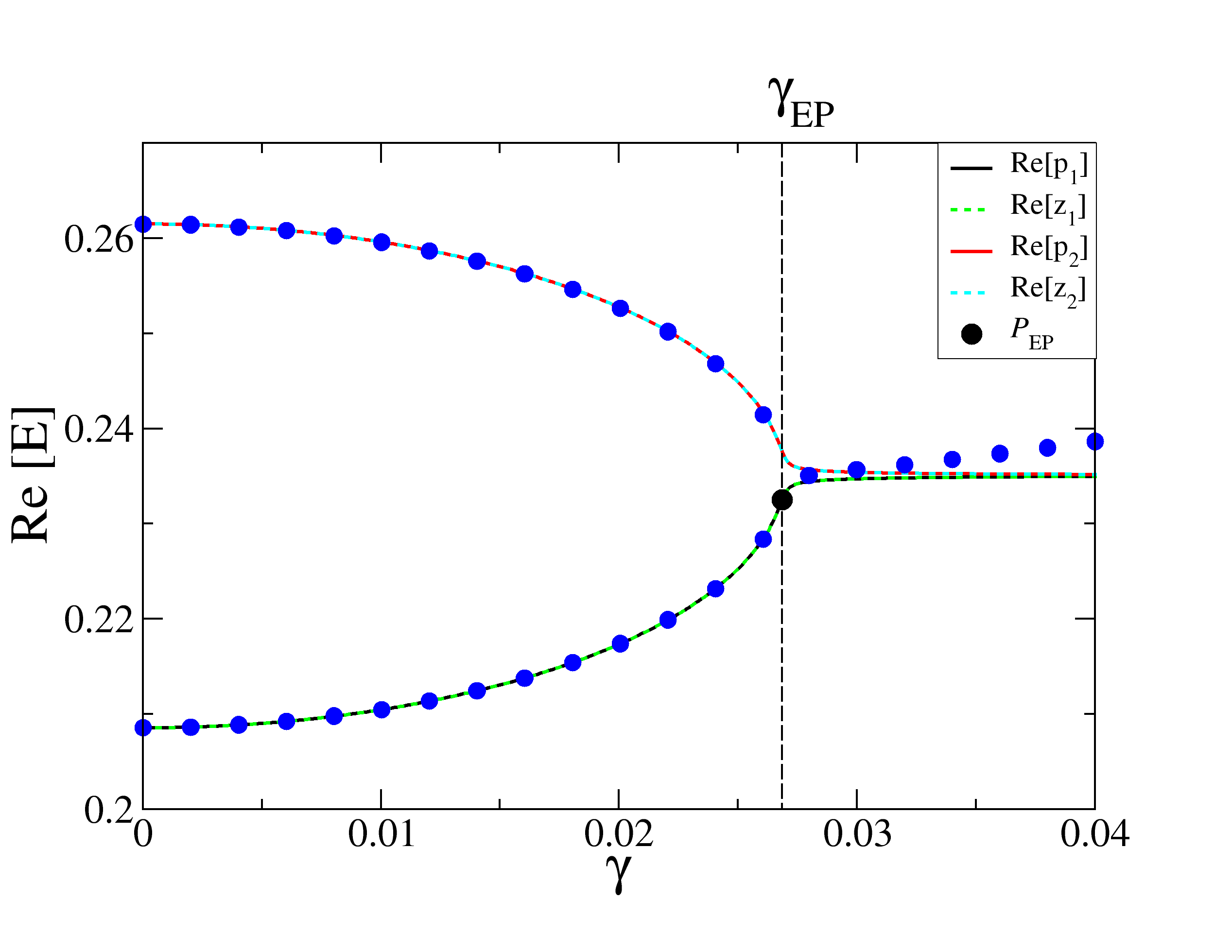}
	\caption{Real parts of the pole and zero trajectories as a function of $\gamma$, together with the resonance-peak positions already reported in Fig.~\ref{fig2:TransmissionSpectrum} (blue circles). The vertical dashed line marks $\gamma_{\textrm{EP}}\approx0.02685$ obtained from resonance-peak tracking (Sec.~\ref{subsec:EP_identification}), and the black point $P_{\textrm{EP}}$ marks the algebraically determined exceptional point (Sec.~\ref{subsec:algebraic-EP}). This confirms visually that the two determinations coincide. The two branches approach each other, change concavity, and settle without crossing, an avoided crossing of energies.}
	\label{fig:poles-real-part}
\end{figure}

First, $\textrm{Re}(p_1)$ and $\textrm{Re}(p_2)$ approach one another, change concavity near $\gamma\approx0.0265$--$0.0269$, and subsequently level off \emph{without crossing} (see Fig.~\ref{fig:poles-real-part})---an avoided crossing of energies, in direct analogy with level repulsion in the Hermitian case. A direct search for the point of closest approach between $p_1$ and $p_2$ finds a finite, non-vanishing minimum separation of $|p_1-p_2|\approx6\times10^{-3}$~eV at $\gamma\approx0.0269$. Within this precision, the two poles of the doublet do not coalesce in the complex plane, and this system does not exhibit a Kato-type double pole at real $\gamma$.

%--------------------------------------------------------------
% FIGURA C (Eduardo): Im(E) vs gamma
%--------------------------------------------------------------
\begin{figure}[htbp]
	\centering
	\includegraphics[width=10cm]{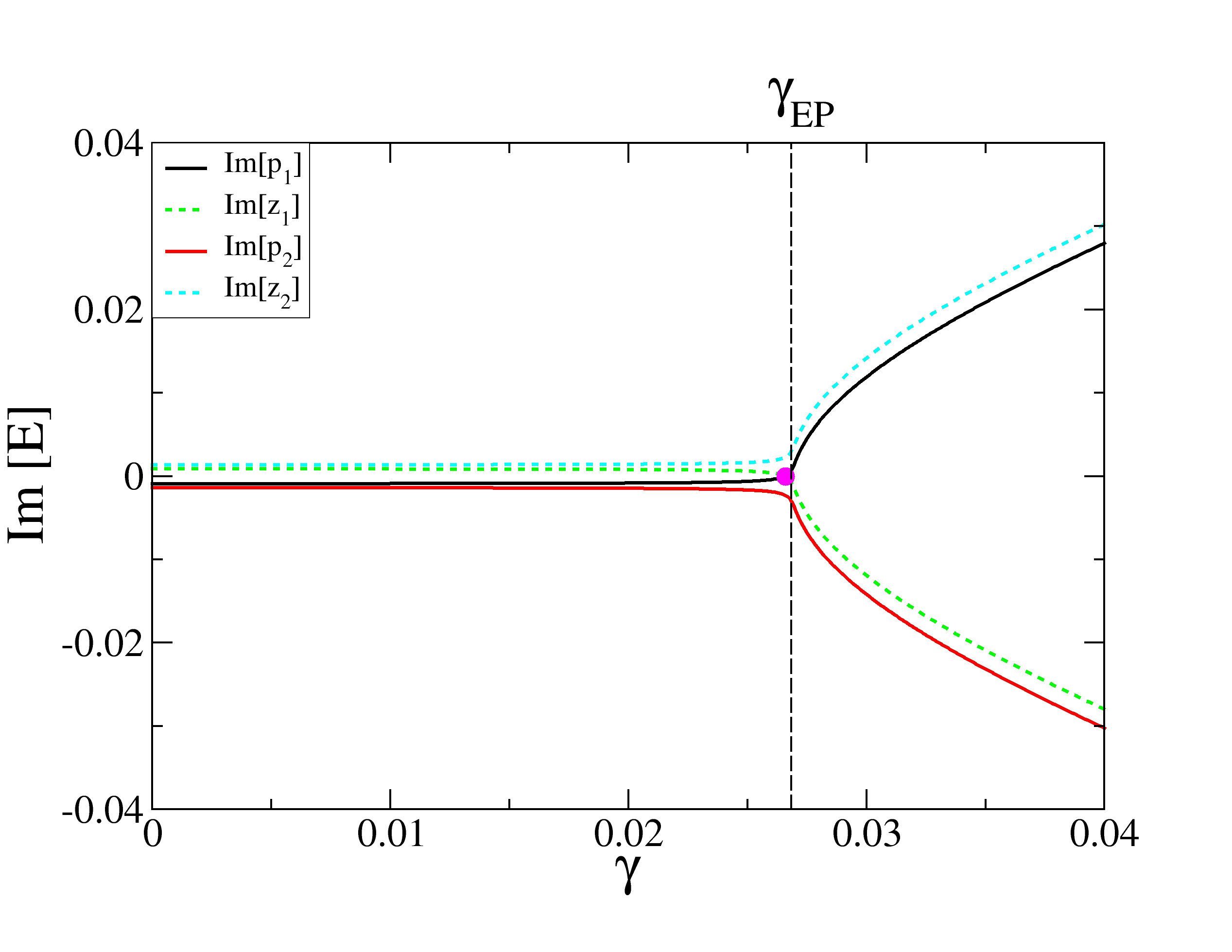}
	\caption{Imaginary parts of the pole and zero trajectories as a function of $\gamma$. Only the lower-energy pair, $p_1$ and $z_1$, crosses the real axis. The higher-energy pair, $p_2$ and $z_2$, diverges monotonically toward more negative and more positive $\textrm{Im}(E)$, respectively, without crossing. The magenta point marks the precise crossing, at $\gamma_{\max}=0.026583$	and $\textrm{Re}(E)=0.230824$~eV.}
	\label{fig:poles-imag-part}
\end{figure}

Second, and in contrast to the real parts, only the lower-energy pole crosses the real axis: $\textrm{Im}(p_1)$ passes through zero together with its Schwarz-conjugate zero $z_1$, both meeting at $\textrm{Re}(E)\approx0.230824$~eV and $\gamma_{max}=0.026583$ (see Fig.~\ref{fig:poles-imag-part}), a value we confirmed independently using two separate numerical approaches. We denote this value by $\gamma_{\max}$, since the vanishing of $\textrm{Im}(p_1)$ is also the condition under which the transmission amplitude and $|\theta_n^{\textrm{Im}}|$ attain their maxima, as discussed below. Beyond this point, $p_1$ moves into the upper half-plane, i.e., the lower-energy resonance transitions from net-decaying to net-amplifying. The higher-energy pole $p_2$ does not cross the real axis anywhere in the explored range: it moves monotonically further into the lower half-plane as $\gamma$ increases, while its mirror zero $z_2$ moves correspondingly further into the upper half-plane, the pair simply separating rather than crossing. This asymmetry between the two members of the doublet traces back to their unequal natural linewidths at $\gamma=0$ ($\Gamma_{w,1}<\Gamma_{w,2}$, evident already in Fig.~\ref{fig:poles-imag-part} at $\gamma=0$). Only the narrower, lower-energy resonance is close enough to the real axis to be driven across it within the explored range of $\gamma$.

The value $\gamma_{\max}$ coincides, to within the precision of both calculations, with two independent real-energy observables, namely the $\gamma$ at which the resonance amplitude reaches its maximum (Sec.~\ref{subsec:EP_identification}) and the inflection point of $G(\gamma)$ (Sec.~\ref{subsec:Subsec33}). This is not coincidental: for a Lorentzian-like resonance, peak transmission scales as $T_{\max}\sim1/\Gamma_{w,1}^2$, so that the resonance-amplitude maximum and the vanishing of $\Gamma_{w,1}=-2\,\textrm{Im}(p_1)$ are the same underlying condition expressed through two different observables. This provides a direct, physically transparent link between the real-energy diagnostics of Secs.~\ref{subsec:EP_identification} and \ref{subsec:Subsec33} and the complex-plane pole dynamics developed here.

\subsection{Convergence of independent diagnostics}
\label{subsec:convergence}

The results of Secs.~\ref{subsec:EP_identification}--\ref{subsec:poles-zeros} should not be read as one phenomenon causing or explaining another: resonance-peak merging, algebraic $S$-matrix eigenvalue degeneracy, and the pole--zero dynamics examined above are formally independent conditions, drawn from established but distinct approaches to non-Hermitian degeneracy in open scattering systems~\cite{Ambichl,Kato,Hernandez2011}. What is notable, and what we report here, is that all three independently place the transition within the same narrow window, $\gamma\approx0.0265$--$0.0269$. This convergence, rather than any single one of these diagnostics in isolation, is the strongest evidence for the location and physical reality of the exceptional point studied in this work.

Within that window, a finer structure is apparent. The vanishing of $\textrm{Im}(p_1)$ defines $\gamma_{\max}=0.026583$ (Sec.~\ref{subsec:poles-zeros}), and two further real-energy observables recover the same point to within $2\times10^{-5}$ in $\gamma$, namely the resonance-amplitude maximum ($\gamma=0.026564$) and the inflection point of $G(\gamma)$ ($\gamma\approx0.026564$, Sec.~\ref{subsec:Subsec33}), despite being extracted from conceptually unrelated quantities. The algebraic exceptional point of Sec.~\ref{subsec:algebraic-EP} lies slightly above them, at $\gamma_{\textrm{EP}}=0.0268500004$. This separation of $\Delta\gamma=\gamma_{\textrm{EP}}-\gamma_{\max}\approx3\times10^{-4}$ is not a numerical discrepancy but a physical one, and is analysed in Sec.~\ref{subsec:Subsec33}.

The absence of a strict pole--pole coalescence in the complex plane (Sec.~\ref{subsec:poles-zeros}) is itself consistent with the codimension-two character of resonance degeneracies~\cite{Hernandez2011}. With only $\gamma$ varied and all geometric parameters of the potential held fixed, the system generically traces a path that passes near, rather than exactly through, the pole--pole degeneracy in the full parameter space, while still crossing the eigenvalue-degeneracy condition of Sec.~\ref{subsec:algebraic-EP} exactly.

%%%%%%%%%%%%%%%%%%%%%%%%%%%%%%%%%%%%%%%%%%%%%%%%%%%%%%%%%%%%%%%%%%%%%%%%%%%%%%%%%%%%%%%%%%%%%%%%%%%%%%%%%%%%%%%%%%%%%%%%%%%%%%%%%%%%%%%%%%%%%%%%%%%%%%%%%%%%%%%%%%%%%%%%%%%%%%%%%%%%%%%%%%%%%%%%%%%%%%%%%%%%%%%%
\section{Scattering eigenphases and the $\mathcal{PT}$-symmetry-breaking transition}
\label{sec:eigenphases-Ggamma}
%%%%%%%%%%%%%%%%%%%%%%%%%%%%%%%%%%%%%%%%%%%%%%%%%%%%%%%%%%%%%%%%%%%%%%%%%%%%%%%%%%%%%%%%%%%%%%%%%%%%%%%%%%%%%%%%%%%%%%%%%%%%%%%%%%%%%%%%%%%%%%%%%%%%%%%%%%%%%%%%%%%%%%%%%%%%%%%%%%%%%%%%%%%%%%%%%%%%%%%%%%%%%%%%

\subsection{Reciprocal pairing and the structure of the eigenphases}
\label{subsec:Subsec31}

In Hermitian systems, flux conservation implies that the scattering matrix is unitary, so that $|\lambda_n|=1$ and all eigenvalues lie on the unit circle~\cite{MelloBook}. The eigenphases are then real and provide a global description of scattering that is independent of any particular resonance representation. In non-Hermitian open systems, by contrast, $S$ is generally non-unitary and its eigenvalues may move away from the unit circle as gain and loss are introduced; the eigenphases $\theta_n = \theta_n(E,\gamma)$ become, in general, complex quantities~\cite{Ambichl, Chong2012}.

For systems with balanced gain and loss, however, the scattering matrix inherits constraints imposed by $\mathcal{PT}$ symmetry. The combined action of parity and time reversal leads to a symmetry relation between the eigenvalues of $S$, such that they occur in reciprocal pairs $\lambda$ and $1/\lambda^*$ ($\lambda_1\lambda_2^*=1$)~\cite{Ambichl}. As a consequence, in the $\mathcal{PT}$-unbroken phase both eigenvalues remain on the unit circle and the eigenphases stay real, despite the non-Hermitian nature of the system. At the exceptional point this protection breaks down: the two eigenvalues coalesce and the eigenphases become degenerate. Beyond the EP, the reciprocal pairing forces one eigenvalue inside and the other outside the unit circle, signaling the onset of the $\mathcal{PT}$-broken phase. Importantly, this entire transition is encoded directly in the eigenphases of $S$, without reference to individual resonances or quasi-bound states.

The scattering matrix formalism allows one to characterize not only transmission and reflection amplitudes, but also the global phase properties of the open system. Diagonalizing $S$ yields two eigenvalues $\lambda_n$ ($n=1,2$), written explicitly as
\begin{equation}
	\begin{bmatrix} \lambda_1 & 0 \\ 0 & \lambda_2 \end{bmatrix}=
	\begin{bmatrix} 
		\frac{1}{2}\left(r+r'+\sqrt{(r-r')^2+4tt'}\right) & 0 \\
		0 & \frac{1}{2}\left(r+r'-\sqrt{(r-r')^2+4tt'}\right)
	\end{bmatrix}.
	\label{ec2:matriz_diagonalizada}
\end{equation}
In the Hermitian limit, where $r=r'$ and $t=t'$, Eq.~\eqref{ec2:matriz_diagonalizada} reduces to $\lambda_{1,2}=r\pm t$, recovering the eigenvalue expressions~\cite{Mares2009}. Each eigenvalue can be expressed in polar form as $\lambda_n = e^{\textrm{i}\theta_n}$, where $\theta_n$ are complex scattering eigenphases. Writing each eigenphase as
\begin{equation}
	\theta_n = \theta_n^{\textrm{Re}} + \textrm{i}\theta_n^{\textrm{Im}},
\end{equation}
the corresponding eigenvalue takes the form
\begin{equation}
	\lambda_n = e^{\textrm{i}\theta_n^{\textrm{Re}}}e^{-\theta_n^{\textrm{Im}}}.
	\label{eq:lambda_decomposition}
\end{equation}
The factor $e^{\textrm{i}\theta_n^{\textrm{Re}}}$ encodes the scattering phase shift, while $e^{-\theta_n^{\textrm{Im}}}$ controls the global amplification or attenuation of the eigenmode. In particular, $\theta_n^{\textrm{Im}}>0$ corresponds to global attenuation $(|\lambda_n|<1)$, whereas $\theta_n^{\textrm{Im}}<0$ corresponds to global amplification $(|\lambda_n|>1)$~\cite{Economou}. Considering these definitions, we can now recall that the reciprocal pairing $\lambda_1\lambda_2^*=1$ separates into two statements of different character: the modulus condition gives $\theta_1^{\textrm{Im}}+\theta_2^{\textrm{Im}}=0$ \emph{exactly}, with no branch ambiguity, since $\theta_n^{\textrm{Im}}=-\ln|\lambda_n|$ is single-valued; the phase condition gives $\theta_1^{\textrm{Re}}-\theta_2^{\textrm{Re}}=0$ only modulo $2\pi$, reflecting the branch ambiguity of $\arg\lambda_n$. This distinction underlies the simplification of $G(\gamma)$ introduced in Sec.~\ref{subsec:Subsec33}, and provides a branch-safe consistency check on the analytical expressions of Appendix~A.

Figure~\ref{fig4:ReThetas} shows the real parts $\theta_n^{\textrm{Re}}$ and Fig.~\ref{fig5:ImThetas} the imaginary parts $\theta_n^{\textrm{Im}}$ of both eigenphases as functions of energy for increasing $\gamma$. For $\gamma=0$ the eigenphases are purely real and each resonance produces the expected $\pi$ phase advance. As $\gamma$ increases, $\theta_n^{\textrm{Im}}$ remains negligible until the system approaches the exceptional point, where a localized structure emerges signaling the onset of gain/loss asymmetry in the scattering eigenmodes. This behavior motivates the introduction of a single integrated quantity that captures the $\mathcal{PT}$ transition globally, without requiring energy-resolved measurements at each $\gamma$.
\begin{figure*}
	\includegraphics[width=12cm]{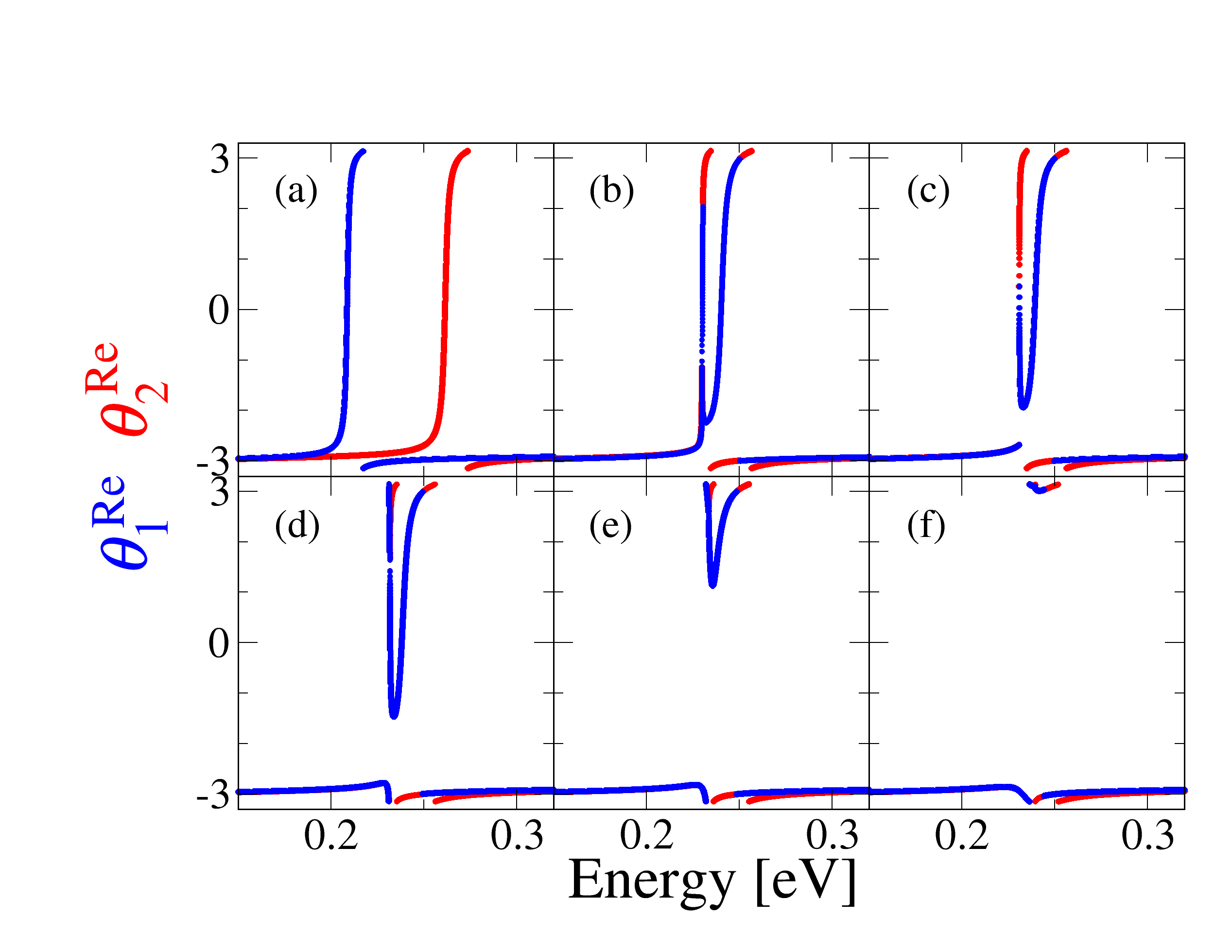}
	\caption{(Color online) Real parts $\theta_{1}^{\textrm{Re}}$ (blue) and $\theta_{2}^{\textrm{Re}}$ (red) of the scattering eigenphases as functions of energy for increasing non-Hermitian parameter $\gamma$. Dimer parameters are the same as in Fig.~\ref{fig2:TransmissionSpectrum}. (a) For $\gamma=0$ the system is Hermitian and the two eigenphases are well separated, each producing a characteristic $\pi$ phase advance associated with a distinct resonance of the doublet. (b) The two phase features approach each other in energy as the resonances begin to overlap. (c) The phase features are nearly degenerate, reflecting the proximity to the exceptional point; the higher-energy phase shift begins to lose resolution. (d) The two eigenphases are on the verge of coalescence, just below $\gamma_{\textrm{EP}}\approx 0.02685$; only a single broadened phase structure is resolvable. Panels (e) and (f) are beyond the EP in the $\mathcal{PT}$-broken phase, the two eigenphases have fully coalesced into a single feature that evolves smoothly with energy, consistent with the merging of the two resonances visible in Fig.~\ref{fig2:TransmissionSpectrum}.}
	\label{fig4:ReThetas}
\end{figure*}
\begin{figure*}
	\includegraphics[width=12cm]{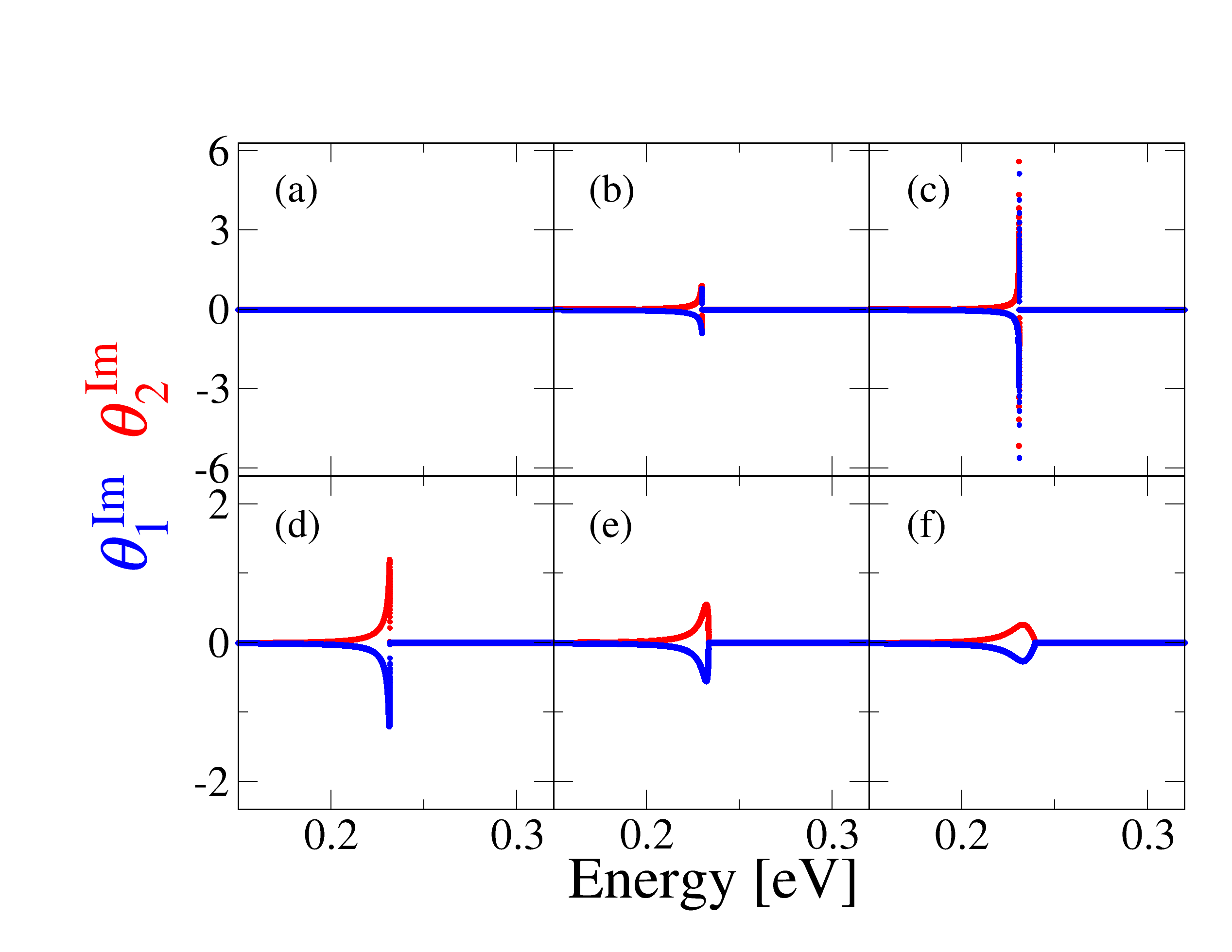}
	\caption{(Color online) Imaginary parts $\theta_{1}^{\textrm{Im}}$ (blue) and $\theta_{2}^{\textrm{Im}}$ (red) of the scattering eigenphases as functions of energy for the same values of $\gamma$ as in Fig.~\ref{fig4:ReThetas}. (a) Both imaginary parts vanish identically, consistent with the Hermitian limit. (b) Finite imaginary parts emerge near the resonance energies with opposite signs, indicating that one eigenmode acquires global gain while the other undergoes global loss. (c) The imaginary parts reach their maximum amplitude, with sharp, large-amplitude structures of opposite sign reflecting the peak gain/loss asymmetry of the scattering eigenmodes. Notably, this maximum occurs at $\gamma_{\max}$, before the exceptional point, consistent with the inflection of $G(\gamma)$ shown in Fig.~\ref{fig6:Integral}. (d) The amplitude of $\theta_{n}^{\textrm{Im}}$ has already begun to decrease even as the system has not yet reached $\gamma_{\textrm{EP}}$, illustrating that the peak gain/loss asymmetry and the coalescence of eigenstates are governed by distinct mechanisms in this open system. Beyond the EP, panels (e) and (f), $\theta_{n}^{\textrm{Im}}$ continues to decrease slowly and the two contributions merge into a single broadened structure, consistent with the saturation of $G(\gamma)$ into a plateau in the $\mathcal{PT}$-broken phase.}
	\label{fig5:ImThetas}
\end{figure*}

\subsection{Integrated imaginary eigenphase as a global diagnostic of the $\mathcal{PT}$-symmetry-breaking transition}
\label{subsec:Subsec33}

The eigenphase analysis of Sec.~\ref{subsec:Subsec31} is a reinterpretation, at the level of a single physical observable, of the same transition already located in Sec.~\ref{sec:EPlocation}. The energy-resolved analysis of Sec.~\ref{subsec:Subsec31} reveals that the imaginary parts of the scattering eigenphases, $\theta_{n}^{\textrm{Im}}(E,\gamma)$, grow progressively as the non-Hermitian parameter approaches the exceptional point. While this behavior provides a detailed picture of the $\mathcal{PT}$-symmetry-breaking transition, its extraction requires resolving the full energy dependence of $\theta_{n}^{\textrm{Im}}$ at each value of $\gamma$---a demanding requirement in experimental settings where phase-resolved measurements are available but spectral resolution is limited. This motivates the introduction of a single scalar quantity that integrates this information into a globally accessible observable.

We define the integrated imaginary scattering eigenphase as
\begin{eqnarray}
	G(\gamma)=\int{\left[\left|\theta_{1}^{\textrm{Im}}(\gamma)\right|+\left|\theta_{2}^{\textrm{Im}}(\gamma)\right|\right]}dE,
	\label{eq:G}
\end{eqnarray}
where the integration is performed over the relevant energy window containing the scattering resonances of interest (in our case, first doublet). By construction, $G(\gamma)$ collapses the full energy-dependent content of both imaginary eigenphases into a single number that tracks the total gain/loss asymmetry accumulated across the scattering spectrum as a function of $\gamma$ alone. Owing to the exact identity $\theta_2^{\textrm{Im}}=-\theta_1^{\textrm{Im}}$ established in Sec.~\ref{subsec:Subsec31}, Eq.~\eqref{eq:G} reduces to $G(\gamma)=2\int|\theta_1^{\textrm{Im}}|\,dE$. However, we retain the symmetric form above to emphasize the physical content of both eigenmodes. In the Hermitian limit, $G(0)=0$ exactly, and $G(\gamma)$ remains negligibly small throughout most of the $\mathcal{PT}$-exact phase, where the eigenvalues are constrained to the unit circle and $\theta_{n}^{\textrm{Im}}\approx0$.

\begin{figure}
	\centering
	\includegraphics[width=10cm]{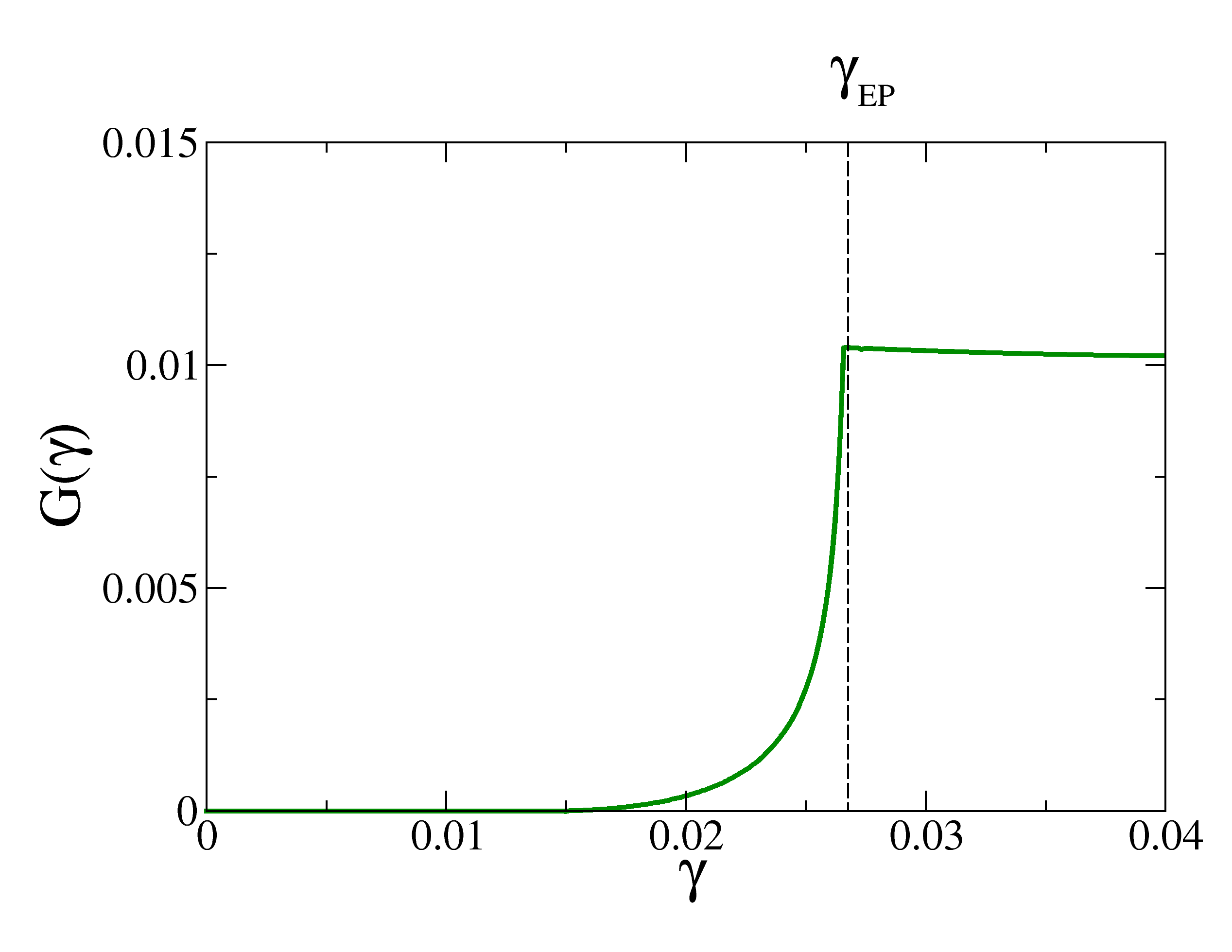}
	\caption{(Color online) Integrated imaginary scattering eigenphase $G(\gamma)$ as a function of the non-Hermitian parameter $\gamma$ for the $\mathcal{PT}$-symmetric dimer. The vertical dashed line marks $\gamma_{\textrm{EP}}\approx 0.02685$, established independently in Sec.~\ref{sec:EPlocation}. Three regimes are visible. For small $\gamma$ the quantity remains negligibly small, consistent with the $\mathcal{PT}$-exact phase. As $\gamma$ increases, $G(\gamma)$ grows with increasing rapidity and reaches a pronounced inflection point at $\gamma\approx 0.026564$, recovering $\gamma_{\max}$ (Sec.~\ref{subsec:poles-zeros}) and coinciding with the maximum amplitude of the transmission resonances shown in Fig.~\ref{fig2:TransmissionSpectrum}(c) and with the maximum of $|\theta_n^{\textrm{Im}}|$ shown in Fig.~\ref{fig5:ImThetas}(c). Beyond $\gamma_{\textrm{EP}}$, $G(\gamma)$ saturates into a plateau, confirming the onset of the $\mathcal{PT}$-broken phase.}
	\label{fig6:Integral}
\end{figure}

The behavior of $G(\gamma)$ is shown in Fig.~\ref{fig6:Integral} for the $\mathcal{PT}$-symmetric dimer studied throughout this work. Three qualitatively distinct regimes are visible. For small $\gamma$, $G(\gamma)\approx 0$, consistent with the near-Hermitian character of the system. As $\gamma$ increases, $G(\gamma)$ begins to grow, slowly at first and then with increasing rapidity, reflecting the progressive development of gain/loss asymmetry in the scattering eigenmodes shown in Fig.~\ref{fig5:ImThetas}. This growth culminates in a pronounced inflection point located at $\gamma\approx 0.026564$, after which $G(\gamma)$ rises more steeply before saturating into a plateau in the $\mathcal{PT}$-broken phase.

A natural question is whether the inflection point of $G(\gamma)$ coincides with the exceptional point located in Sec.~\ref{sec:EPlocation}. The answer is instructive: it does not exactly, but the discrepancy is itself physically meaningful. Comparing with the transmission spectra of Fig.~\ref{fig2:TransmissionSpectrum}, one finds that the inflection point locates $\gamma_{\max}$ instead, the $\gamma$ at which the resonance amplitudes reach their maximum (panel (c)). This is the point at which the imaginary parts of the resonance energies are minimal, i.e., the resonances are sharpest and the system is closest to a lasing condition. The exceptional point itself, established in Sec.~\ref{sec:EPlocation} at $\gamma_{\textrm{EP}}\approx0.02685$ and corroborated by two further independent analyses, occurs slightly later, after the amplitudes have already begun to decrease. In an open non-Hermitian system, these two events are governed by independent conditions on $\textrm{Re}[E]$ and $\textrm{Im}[E]$ respectively, and there is no general reason for them to coincide. Their separation, $\Delta\gamma=\gamma_{\textrm{EP}}-\gamma_{\max}\approx0.0003$, is a direct fingerprint of the open character of the system and its finite coupling to the continuum.

Despite this distinction, $G(\gamma)$ retains clear practical value as an experimental diagnostic, constituting a fourth, methodologically distinct route to the same physics established in Sec.~\ref{sec:EPlocation}: unlike resonance-peak tracking, algebraic eigenvalue degeneracy, or complex pole tracking, $G(\gamma)$ requires only phase-resolved measurements of the scattering matrix eigenvalues, integrated over an energy window, without the need to solve any algebraic condition at a specific point or to track spectral features individually. The inflection point of $G(\gamma)$ provides a lower bound on $\gamma_\textrm{EP}$ that is sharp, unambiguous, and directly readable from the data. Furthermore, as the system enters the $\mathcal{PT}$-broken phase, the saturation of $G(\gamma)$ into a plateau provides an independent confirmation that the transition has occurred---a feature with no counterpart in amplitude-based observables.

%%%%%%%%%%%%%%%%%%%%%%%%%%%%%%%%%%%%%%%%%%%%%%%%%%%%%%%%%%%%%%%%%%%%%%%%%%%%%%%%%%%%%%%%%%%%%%%%%%%%%%%%%%%%%%%%%%%%%%%%%%%%%%%%%%%%%%%%%%%%%%%%%%%%%%%%%%%%%%%%%%%%%%%%%%%%%%%%%%%%%%%%%%%%%%%%%%%%%%%%%%%%%%%%

\section{Conclusions}
\label{sec:conclusions}

We have demonstrated that the eigenphases of the scattering matrix provide direct and experimentally accessible signatures of exceptional points in open $\mathcal{PT}$-symmetric systems. Using an analytically tractable model of a one-dimensional $\mathcal{PT}$-symmetric quantum dimer coupled to continuum leads, we located the exceptional point at $\gamma_{\textrm{EP}}\approx0.02685$ through the algebraic degeneracy condition of the $S$-matrix eigenvalues, and corroborated it by resonance-peak tracking and by the dynamics of the complex poles and zeros of $S(E)$, whose convergence on the same narrow window establishes its location and physical reality. We then tracked the evolution of the scattering eigenphases across the $\mathcal{PT}$-exact to $\mathcal{PT}$-broken transition and identified two complementary observables that encode this transition at different levels of resolution.

At the energy-resolved level, the real parts of the eigenphases exhibit the progressive coalescence of the characteristic $\pi$ phase advances associated with each resonance of the doublet, while their imaginary parts develop localized structures of opposite sign that signal the emergence of gain/loss asymmetry in the scattering eigenmodes. Crucially, the amplitude of these imaginary parts reaches a maximum and begins to decrease before the exceptional point is reached---a behavior that reflects the open character of the system, where the condition for maximum gain/loss asymmetry and the condition for eigenstate coalescence are governed by independent mechanisms and need not coincide.

At the global level, the integrated imaginary scattering eigenphase $G(\gamma)$ condenses this information into a single scalar quantity that is directly readable from phase-resolved scattering measurements. $G(\gamma)$ remains negligibly small throughout most of the $\mathcal{PT}$-exact phase and undergoes a sharp transition---marked by a pronounced inflection point followed by rapid growth and subsequent saturation into a plateau---in the vicinity of the exceptional point. The inflection point of $G(\gamma)$ coincides with the maximum amplitude of the transmission resonances and provides a robust lower bound on $\gamma_{\textrm{EP}}$, while the onset of the plateau independently confirms that the $\mathcal{PT}$-broken phase has been entered. Together, these two features bracket the exceptional point from below and above, offering a characterization of the $\mathcal{PT}$ transition that does not require resolving individual resonance peaks---a significant practical advantage in noisy or multimode environments where amplitude-based methods become unreliable.

Because our analysis is formulated entirely at the level of the scattering matrix, without recourse to system-specific Hamiltonian representations or tight-binding approximations, the proposed signatures are not restricted to the quantum mechanical setting considered here. They are directly applicable to any wave system describable by an $S$-matrix---including photonic, acoustic, and microwave platforms---where phase-resolved measurements are already performed using vector network analyzers. We therefore expect $G(\gamma)$ to serve as a practical and universal diagnostic tool for exceptional points in realistic open non-Hermitian systems, complementing and extending the reach of existing amplitude-based approaches.

%%%%%%%%%%%%%%%%%%%%%%%%%%%%%%%%%%%%%%%%%%%%%%%%%%%%%%%%%%%%%%%%%%%%%%%%%%%%%%%%%%%%%%%%%%%%%%%%%%%%%%%%%%%%%%%%%%%%%%%%%%%%%%%%%%%%%%%%%%%%%%%%%%%%%%%%%%%%%%%%%%%%%%%%%%%%%%%%%%%%%%%%%%%%%%%%%%%%%%%%%%%%%%%%

\begin{acknowledgments}
	J. C-G and J.~E. T-Z thank financial support from CONAHCyT. V. D-R is grateful with M. Dom\'inguez-Tello for his encouragement. %Victor is grateful to Mario for always encouraging me not to give up in his own way.
	The authors thank useful comments and discussions of E. Casta\~no, M. Mart\'inez-Mares, A.~A.~Fernández-Marín and J.~A. Franco-Villafa\~ne.
\end{acknowledgments}
	
%%%%%%%%%%%%%%%%%%%%%%%%%%%%%%%%%%%%%%%%%%%%%%%%%%%%%%%%%%%%%%%%%%%%%%%%%%%%%%%%%%%%%%%%%
%%%%%%%%%%%%%%%%%%%%%%%%%%%%%%%%%%%%%%%%%%%%%%%%%%%%%%%%%%%%%%%%%%%%%%%%%%%%%%%%%%%%%%%%%
%%%%%%%%%%%%%%%%%%%%%%%%%%%%%%%%%%%%%%%%%%%%%%%%%%%%%%%%%%%%%%%%%%%%%%%%%%%%%%%%%%%%%%%%%
	
%\appendix
\section{Appendix}

\subsection{Analytical $S$ matrix of the $\mathcal{PT}$-symmetric dimer}

In this Appendix we collect the analytical expressions required to compute the scattering matrix of the $\mathcal{PT}$-symmetric quantum dimer introduced in Sec.~\ref{subsec:model}. These results are provided for completeness and reproducibility, while the main text focuses on the physical properties of the scattering eigenphases and their evolution across the $\mathcal{PT}$-symmetry-breaking transition.

The potential profile consists of seven regions, as illustrated in Fig.~\ref{fig1:Diagram}. Regions I and VII correspond to free leads with $V(x)=0$. Regions II, IV, and VI are real potential barriers of height $V_{\textrm{b}}$, while regions III and V contain complex potentials $V' \mp \textrm{i}\gamma$, representing spatially separated loss and gain, respectively. The energy range considered satisfies $V_{\textrm{b}} > E > V'$, ensuring evanescent waves in the real barriers and resonant behavior in the non-Hermitian regions.

\subsection*{Wave numbers and stationary solutions}

In each region the stationary solutions of the Schr\"odinger equation are written as superpositions of plane waves traveling to the left and right. The corresponding wave numbers are
\begin{equation}
	k_1 = k_7 = \sqrt{\frac{2mE}{\hbar^2}}, \qquad
	k_2 = k_4 = k_6 = \sqrt{\frac{2m(E - V_{\textrm{b}})}{\hbar^2}},
\end{equation}
\begin{equation}
	k = \sqrt{\frac{2mE}{\hbar^2}} \quad
	k_{\textrm{b}} = \sqrt{\frac{2m(E - V_{\textrm{b}})}{\hbar^2}}\quad
	k_- = \sqrt{\frac{2m}{\hbar^2}\left[E-(V' - \textrm{i}\gamma)\right]} \quad
	k_+ = \sqrt{\frac{2m}{\hbar^2}\left[E-(V' + \textrm{i}\gamma)\right]}
\end{equation}
corresponding for the free and real-barrier regions, and
\begin{equation}
	k_3 = \sqrt{\frac{2m}{\hbar^2}\left[E-(V' - \textrm{i}\gamma)\right]}, \qquad
	k_5 = \sqrt{\frac{2m}{\hbar^2}\left[E-(V' + \textrm{i}\gamma)\right]},
\end{equation}
for the loss and gain regions, respectively. Since the energy range satisfies \(V_{\textrm{b}} > E > V'\), the wave numbers in regions II, IV, and VI are purely imaginary and can be written as
\begin{equation}
	k_2 = k_4 = k_6 = \textrm{i}q, \qquad
	q = \sqrt{\frac{2m(V_{\textrm{b}} - E)}{\hbar^2}}.
\end{equation}
The complex wave numbers in the non-Hermitian regions can be decomposed as
\begin{eqnarray}
	k_3 &=& K + \textrm{i}\Gamma, \\
	k_5 &=& K - \textrm{i}\Gamma ,
\end{eqnarray}
with
\begin{eqnarray}
	K &=& \sqrt{\frac{m}{\hbar^2}\left[(E - V') + \sqrt{(E - V')^2 + \gamma^2}\right]}, \\
	\Gamma &=& \sqrt{\frac{m}{\hbar^2}\left[\sqrt{(E - V')^2 + \gamma^2} - (E - V')\right]},
\end{eqnarray}
where both \(K\) and \(\Gamma\) are positive real quantities. The parameter \(\Gamma\) controls exponential attenuation (loss) or amplification (gain) in regions III and V, respectively.

The stationary wave functions in the seven regions are therefore given by
\begin{equation}
	\begin{aligned}
		\psi_{1}(x) &= a_1 e^{\textrm{i}k_1 x} + b_1 e^{-\textrm{i}k_1 x}, \\
		\psi_{2}(x) &= b_2 e^{-q x} + a_2 e^{q x}, \\
		\psi_{3}(x) &= b_3 e^{\textrm{i}K x}e^{-\Gamma x} + a_3 e^{-\textrm{i}K x}e^{+\Gamma x}, \\
		\psi_{4}(x) &= b_4 e^{-q x} + a_4 e^{q x}, \\
		\psi_{5}(x) &= b_5 e^{\textrm{i}K x}e^{+\Gamma x} + a_5 e^{-\textrm{i}K x}e^{-\Gamma x}, \\
		\psi_{6}(x) &= b_6 e^{-q x} + a_6 e^{q x}, \\
		\psi_{7}(x) &= b_7 e^{\textrm{i}k_1 x} + a_7 e^{-\textrm{i}k_1 x}.
	\end{aligned}
	\label{eq:Psi}
\end{equation}
The explicit sign of \(\Gamma\) in regions III and V highlights the presence of loss and gain, respectively. For example, the term
\(b_3 e^{\textrm{i}Kx}e^{-\Gamma x}\) represents a right-propagating wave that decays along the loss region, while \(b_5 e^{\textrm{i}Kx}e^{+\Gamma x}\) corresponds to amplification in the gain region.

\subsection*{Construction of the scattering matrix}

The scattering matrix is obtained by imposing continuity of the wave functions and their first derivatives at all interfaces between adjacent regions. This procedure relates the outgoing amplitudes $(b_1, b_7)$ to the incoming ones $(a_1, a_7)$ and yields the $2\times2$ scattering matrix
\begin{equation}
S=\begin{bmatrix}
	r & t'\\
	t & r'    
\end{bmatrix},
\end{equation}
where \(r\) and \(t\) (\(r'\) and \(t'\)) denote reflection and transmission amplitudes for incidence from the left (right), respectively. The explicit analytical expressions for the reflection and transmission amplitudes are given by
\begin{eqnarray}
	%%%%%%%%%%%%%%%%% primer t\'ermino %%%%%%%%%%%%%%%%%%%%%
	r&=& r_{12}                                                  
	%%%%%%%%%%%%%%%%% segundo t\'ermino %%%%%%%%%%%%%%%%%%%%%
	+t'_{12}\; \frac{1}{\textrm{e}^{-2\textrm{i}k_1b}-r'_{12}\;r_{23}} \; r_{23}  \;t_{12}  \nonumber\\            
	%%%%%%%%%%%%%%%%% tercer t\'ermino %%%%%%%%%%%%%%%%%%%%%
	&+&t'_{12}\; 
	\frac{\textrm{e}^{-\textrm{i}k_{1}b}}{\textrm{e}^{-2\textrm{i}k_1b}-r'_{12}\;r_{23}}\;  t'_{23} 
	\;\frac{1}{\textrm{e}^{-2\textrm{i}k_{2}a}-r'_{1-3}\;r_{34}}\;r_{34}
	\;t_{23}\; \frac{\textrm{e}^{-\textrm{i}k_1b}}{\textrm{e}^{-2\textrm{i}k_1b}-r'_{12}\;r_{23}} \;t_{12} \nonumber\\
	%%%%%%%%%%%%%%%%% cuarto t\'ermino %%%%%%%%%%%%%%%%%%%%%
	&+&  t'_{12}\frac{\textrm{e}^{-\textrm{i}k_{1}b}}{\textrm{e}^{-2\textrm{i}k_1b}-r'_{12}\;r_{23}}\; t'_{23} 
	\frac{\textrm{e}^{-\textrm{i}k_{2}a}}{\textrm{e}^{-2\textrm{i}k_{2}a}-r'_{1-3}\;r_{34}}
	t'_{34}	\frac{1}{\textrm{e}^{-2\textrm{i}k_{1}b/2}-r'_{1-4}\;r_{45}}\;r_{45} 
	\;  t_{34}\nonumber\\
%	\label{eq2:reflection}\\
	&\times&  
	\frac{\textrm{e}^{-\textrm{i}k_{2}a}}{\textrm{e}^{-2\textrm{i}k_{2}a}-r'_{1-3}\;r_{34}}
	\;t_{23}\frac{\textrm{e}^{-\textrm{i}k_1b}}{\textrm{e}^{-2\textrm{i}k_1b}-r'_{12}\;r_{23}}   
	\;t_{12}\nonumber\\
	%%%%%%%%%%%%%%%%% quinto t\'ermino %%%%%%%%%%%%%%%%%%%%%
	&+&t'_{12}\;\frac{\textrm{e}^{-\textrm{i}k_{1}b}}{\textrm{e}^{-2\textrm{i}k_1b}-r'_{12}\;r_{23}}\;t'_{23}\;\frac{\textrm{e}^{-\textrm{i}k_{2}a}}{\textrm{e}^{-2\textrm{i}k_{2}a}-r'_{1-3}\;r_{34}}	\;t'_{34}\;\frac{\textrm{e}^{-\textrm{i}k_{1}b/2}}{\textrm{e}^{-2\textrm{i}k_{1}b/2}-r'_{1-4}\;r_{45}} \;t'_{45}\; 
	\nonumber\\
	&\times& 
	\frac{1}{\textrm{e}^{-2\textrm{i}k_{2}a}-r'_{1-5}\;r_{56}}  \;r_{56} t_{45}\frac{\textrm{e}^{-\textrm{i}k_{1}b/2}}{\textrm{e}^{-2\textrm{i}k_{1}b/2}-r'_{1-4}\;r_{45}}\; t_{34}\;\frac{\textrm{e}^{-\textrm{i}k_{2}a}}{\textrm{e}^{-2\textrm{i}k_{2}a}-r'_{1-3}\;r_{34}} 
	\;t_{23}\;	\nonumber\\
	&\times& 
	\frac{\textrm{e}^{-\textrm{i}k_1b}}{\textrm{e}^{-2\textrm{i}k_1b}-r'_{12}\;r_{23}} \;t_{12}\\
	\label{eq:reflectiona}
	%%%%%%%%%%%%%%%%% sexto t\'ermino %%%%%%%%%%%%%%%%%%%%%
	&+&t'_{12}\;\frac{\textrm{e}^{-\textrm{i}k_{1}b}}{\textrm{e}^{-2\textrm{i}k_1b}-r'_{12}\;r_{23}}\;t'_{23}\;\frac{\textrm{e}^{-\textrm{i}k_{2}a}}{
		\textrm{e}^{-2\textrm{i}k_{2}a}-r'_{1-3}\;r_{34}}
	\;t'_{34} 
	\;\frac{\textrm{e}^{-\textrm{i}k_{1}b/2}}{\textrm{e}^{-2\textrm{i}k_{1}b/2}-r'_{1-4}\;r_{45}}\; t'_{45} \;\nonumber\\
	&\times&\;\frac{\textrm{e}^{-\textrm{i}k_{2}a}}{\textrm{e}^{-2\textrm{i}k_{2}a}-r'_{1-5}\;r_{56}} t'_{56} 
	\frac{1}{\textrm{e}^{-2\textrm{i}k_{1}b}-r'_{1-6}\;r_{67}}\;r_{67}\; t_{56}\frac{\textrm{e}^{-\textrm{i}k_{2}a}}{\textrm{e}^{-2\textrm{i}k_{2}a}-r'_{1-5}\;r_{56}} 
	\;t_{45} \nonumber\\
	&\times&
	\frac{\textrm{e}^{-\textrm{i}k_{1}b/2}}{\textrm{e}^{-2\textrm{i}k_{1}b/2}-r'_{1-4}\;r_{45}}\;t_{34}\frac{\textrm{e}^{-\textrm{i}k_{2}a}}{\textrm{e}^{-2\textrm{i}k_{2}a}-r'_{1-3}\;r_{34}}	
	\;t_{23}\frac{\textrm{e}^{-\textrm{i}k_1b}}{\textrm{e}^{-2\textrm{i}k_1b}-r'_{12}\;r_{23}}\;t_{12}\nonumber,
\end{eqnarray}
\begin{eqnarray}
	t&=&t_{67}\frac{\textrm{e}^{-\textrm{i}k_{1}b}}{\textrm{e}^{-2\textrm{i}k_{1}b}-r'_{1-5}r_{67}}t_{56} 
	\frac{\textrm{e}^{-\textrm{i}k_{2}a}}{\textrm{e}^{-2\textrm{i}k_{2}a}-r'_{1-4}r_{56}}	
	t_{45} \frac{\textrm{e}^{-\textrm{i}k_{1}b/2}}{\textrm{e}^{-2\textrm{i}k_{1}b/2}-r'_{1-3}r_{45}}t_{34}\frac{\textrm{e}^{-\textrm{i}k_{2}a}}{\textrm{e}^{-2\textrm{i}k_{2}a}-r'_{1-2}r_{34}} t_{23}\nonumber\\
	&\times&
	\frac{\textrm{e}^{-\textrm{i}k_1b}}{\textrm{e}^{-2\textrm{i}k_1b}-r'_{12}r_{23}}  t_{12},
	\label{eq:transmissiona}
\end{eqnarray}
\begin{eqnarray}
	%%%%%%%%%%%%%%%%%%%%%%%%%% primer t\'ermino %%%%%%%%%%%%%%%%%%%%%%%%%%%
	r'&=&r'_{67}
	%%%%%%%%%%%%%%%%%%%%%%%%%% segundo t\'ermino %%%%%%%%%%%%%%%%%%%%%%%%%%%
	+t_{67} \; \frac{1}{
		\textrm{e}^{-2\textrm{i}k_{1}b}-r'_{56}r_{67}} \; 
	r'_{56} \; t'_{67}\nonumber\\
	%%%%%%%%%%%%%%%%%%%%%%%%%% tercer t\'ermino %%%%%%%%%%%%%%%%%%%%%%%%%%%
	&+&  t_{67} \; 
	\frac{\textrm{e}^{-\textrm{i}k_{1}b}}{\textrm{e}^{-2\textrm{i}k_{1}b}
		-r'_{56}r_{67}}  \; t_{56}  \; 
	\frac{1}{\textrm{e}^{-2\textrm{i}k_{2}a}-r'_{45}r_{7-5}}  \; r'_{45}  
	\; t'_{56} \;  
	\frac{\textrm{e}^{-\textrm{i}k_{1}b}}{\textrm{e}^{-2\textrm{i}k_{1}b}-r_
		{67}r'_{56}}  \;  t'_{67}\nonumber\\
	%%%%%%%%%%%%%%%%%%%%%%%%%% cuarto t\'ermino %%%%%%%%%%%%%%%%%%%%%%%%%%%
	&+&t_{67}  
	\frac{\textrm{e}^{-\textrm{i}k_{1}b}}{\textrm{e}^{-2\textrm{i}k_{1}b}
		-r'_{56}r_{67}} \; t_{56} \; 
	\frac{\textrm{e}^{-\textrm{i}k_{2}a}}{\textrm{e}^{-2\textrm{i}k_{2}a}
		-r'_{45}r_{7-5}}
	\; t_{45}
	\frac{1}{\textrm{e}^{-2\textrm{i}k_{1}b/2}-r'_{34}r_{7-4}} \;  r'_{34} \; 
	t'_{45}\nonumber\\
	&\times&   
	\frac{\textrm{e}^{-\textrm{i}k_{2}a}}{\textrm{e}^{-2\textrm{i}k_{2}a}
		-r'_{45}r_{7-5}}
	\; t'_{56}
	\frac{\textrm{e}^{-\textrm{i}k_{1}b}}{\textrm{e}^{-2\textrm{i}k_{1}b}
		-r'_{56}r_{67}} \;  t'_{67}\nonumber\\ 
	%%%%%%%%%%%%%%%%%%%%%%%%%% quinto t\'ermino %%%%%%%%%%%%%%%%%%%%%%%%%%%
	&+&t_{67} 
	\frac{\textrm{e}^{-\textrm{i}k_{1}b}}{\textrm{e}^{-2\textrm{i}k_{1}b}
		-r'_{56}r_{67}} \; t_{56} \; 
	\frac{\textrm{e}^{-\textrm{i}k_{2}a}}{\textrm{e}^{-2\textrm{i}k_{2}a}
		-r'_{45}r_{7-5}}
	t_{45} \;   
	\frac{\textrm{e}^{-\textrm{i}k_{1}b/2}}{\textrm{e}^{-2\textrm{i}k_{1}b/2}
		-r'_{34}r_{7-4}} \;  t_{34}\nonumber\\
	&\times&  \frac{1}{\textrm{e}^{-2\textrm{i}k_{2}a}-r'_{23}r_{7-3}}  \; r'_{23} \; t'_{34}  
	\frac{\textrm{e}^{-\textrm{i}k_{1}b/2}}{\textrm{e}^{-2\textrm{i}k_{1}b/2}
		-r'_{34}r_{7-4}} \; t'_{45} \; 
	\frac{\textrm{e}^{-\textrm{i}k_{2}a}}{\textrm{e}^{-2\textrm{i}k_{2}a}
		-r'_{45}r_{7-5}}
	t'_{56} \;\nonumber\\
	&\times&    
	\frac{\textrm{e}^{-\textrm{i}k_{1}b}}{\textrm{e}^{-2\textrm{i}k_{1}b}
		-r'_{56}r_{67}}  \; t'_{67}
	\label{eq:reflectionap}\\ 
	%%%%%%%%%%%%%%%%%%%%%%%%%% sexto t\'ermino %%%%%%%%%%%%%%%%%%%%%%%%%%%
	&+&t_{67} \; 
	\frac{\textrm{e}^{-\textrm{i}k_{1}b}}{\textrm{e}^{-2\textrm{i}k_{1}b}
		-r'_{56}r_{67}} \; t_{56} \; 
	\frac{\textrm{e}^{-\textrm{i}k_{2}a}}{\textrm{e}^{-2\textrm{i}k_{2}a}
		-r'_{45} r_{7-5}}
	\; t_{45} \;  
	\frac{\textrm{e}^{-\textrm{i}k_{1}b/2}}{\textrm{e}^{-2\textrm{i}k_{1}b/2}
		-r'_{34}r_{7-4}}  \; t_{34} \nonumber\\
	&\times&  
	\frac{\textrm{e}^{-\textrm{i}k_{2}a}}{\textrm{e}^{-2\textrm{i}k_{2}a}-r'_{23}r_{7-3}}  
	\; t_{23} \;  
	\frac{1}{\textrm{e}^{-2\textrm{i}k_{1}b}-r'_{12}r_{7-2}} \; r'_{12} \;	t'_{23} \; 
	\frac{\textrm{e}^{-\textrm{i}k_{2}a}}{\textrm{e}^{-2\textrm{i}k_{2}a}
		-r'_{23}r_{7-3}}
	\; t'_{34}\nonumber\\
	&\times&    
	\frac{\textrm{e}^{-\textrm{i}k_{1}b/2}}{\textrm{e}^{-2\textrm{i}k_{1}b/2}
		-r'_{34}r_{7-4}} \; t'_{45} \;
	\frac{\textrm{e}^{-\textrm{i}k_{2}a}}{\textrm{e}^{-2\textrm{i}k_{2}a}
		-r'_{45}r_{7-5}}
	t'_{56} \;  
	\frac{\textrm{e}^{-\textrm{i}k_{1}b}}{\textrm{e}^{-2\textrm{i}k_{1}b}
		-r'_{56}r_{67}}  \; t'_{67}\nonumber,
\end{eqnarray}
and
\begin{eqnarray}
	t'&=&t'_{12}
	\frac{\textrm{e}^{-\textrm{i}k_{1}b}}{\textrm{e}^{-2\textrm{i}k_{1}b}
		-r'_{12} r_{7-2}} \; t'_{23} 
	\frac{\textrm{e}^{-\textrm{i}k_{2}a}}{\textrm{e}^{-2\textrm{i}k_{2}a}
		-r'_{23} r_{7-3}}
	\;t'_{34}
	\frac{\textrm{e}^{-\textrm{i}k_{1}b/2}}{\textrm{e}^{-2\textrm{i}k_{1}b/2}
		-r'_{34} r_{7-4}} \;t'_{45}
	\frac{\textrm{e}^{-\textrm{i}k_{2}a}}{\textrm{e}^{-2\textrm{i}k_{2}a}
		-r'_{45} \; r_{7-5}}
	t'_{56}\nonumber\\
	&\times&  \frac{\textrm{e}^{-\textrm{i}k_{1}b}}{\textrm{e}^{-2\textrm{i}k_{1}b}
		-r'_{56} \; r_{67}} \; t'_{67}.
	\label{eq:transmissionap}
\end{eqnarray}
where the notation for intermediate reflection and transmission coefficients follows the conventions detailed in Appendix~B. These expressions are exact and fully account for multiple scattering processes inside the $\mathcal{PT}$-symmetric dimer.

Although algebraically involved, the above expressions are fully analytical and valid for any one-dimensional wave system describable by a scattering matrix, independently of the specific physical realization\cite{Robledo2020}. In the main text we focus on the spectral properties of $S$ and, in particular, on the behavior of its eigenphases, which provide a direct and robust signature of exceptional points in open $\mathcal{PT}$-symmetric systems.

\subsection{$S$ matrix building process}

The first notation uses two joined subscripts. For example, in $r'_{\eta \delta}$ we refer to the reflection that occurs between the adjacent potentials $\eta$ and $\delta $ when waves incide from the right. The scattering matrix for this case is given by
\begin{equation}
	S_{\delta \eta}=\begin{bmatrix}
		r_{\delta \eta}  & t'_{\delta \eta} \\
		t_{\delta \eta}     & r'_{\delta \eta}
	\end{bmatrix}
	=\begin{bmatrix}
		\frac{k_{\delta}-k_{\eta}}{k_{\delta}+k_{\eta}}  &  
		\frac{2k_{\eta}}{k_{\delta}+k_{\eta}}\\
		\frac{2k_{\delta}}{k_{\delta}+k_{\eta}}    & 
		-\frac{k_{\delta}-k_{\eta}}{k_{\delta}+k_{\eta}}
	\end{bmatrix}  \label{ec2:matriz_scattering_delta_eta},
\end{equation}
where $k_\delta$ and $k_\eta$ are the wave numbers of the region $\delta$ and $\eta$, respectively, while $S_{\delta\eta}$ is the scattering matrix between the $\delta$ and $\eta$ regions which also represents the scatterer between those regions.

The second notation uses two subscripts separated by a hyphen. Combining the scattering matrices $S_{\delta\eta}$ and $S_{\eta\omega}$ gives $S_{\delta-\omega}$. For example, $t_{\delta-\omega}$ refers to the transmission amplitude generated when the waves are transmitted from $\delta$ region to the non-contiguous $\omega$ region when incident waves arrive from the left. Part of the incident wave from the left on the scatterer $S_{\delta \eta}$ is reflected back into the $\delta$ region and the other part is transmitted towards the $\eta$ region. The part that is transmitted travels to the scatterer $S_{\eta\omega}$, gaining a phase in this path, in which a fraction of this wave is reflected in the $\eta$ region and the other one is transmitted towards the $\omega$ region. The wave that was reflected travels again to the scatterer $S_{\delta\eta}$ in which a part of that wave is reflected again into the $\eta$ region and the other part is transmitted to the $\delta$ region. This process of multiple reflections in the region $\eta$ is repeated infinitely. On the one hand, the sum of all the waves in the $\omega$ region is a geometric series which results in $t_{\delta-\omega}$. On the other hand, the sum of the waves in the $\delta$ region, which is also a geometric series, gives $r_{\delta-\omega}$. These two elements are the transmission and reflection amplitudes when waves are incident from the left of the matrix
\begin{eqnarray}
	S_{\delta-\omega}&=&\begin{bmatrix}
		r_{\delta-\omega}	& t'_{\delta-\omega}   \\
		t_{\delta-\omega}	&  r'_{\delta-\omega}
	\end{bmatrix}\nonumber\\
	&=&\begin{bmatrix}
		r_{\delta\eta} +t'_{\delta 
			\eta}\frac{1}{\textrm{e}^{-2\textrm{i}k_{\eta}d} -r'_{\delta 
				\eta}r_{\eta \omega}}r_{\eta \omega}t_{\delta \eta} & t'_{\delta 
			\eta}\frac{\textrm{e}^{-\textrm{i}k_{\eta}d}}{\textrm{e}^{-2\textrm{i}k_{\eta}d} -r'_{\delta \eta}r_{\eta 
				\omega}}t'_{\eta \omega} \\
		t_{\eta \omega}\frac{\textrm{e}^{-\textrm{i}k_{\eta}d}}{\textrm{e}^{-2\textrm{i}k_{\eta}d} -r'_{\delta \eta}r_{\eta\omega}}t_{\delta \eta} & r'_{\eta \omega} +t_{\eta 
			\omega}\frac{1}{\textrm{e}^{-2\textrm{i}k_{\eta}d} -r'_{\delta 
				\eta}r_{\eta \omega}}r'_{\delta \eta}t'_{\eta \omega} 
	\end{bmatrix}.\nonumber
	\label{ec2:combinacion matrices}
\end{eqnarray}
This procedure is also valid if it is necessary to build the system from the right since $S=S_{1-7}=S_{7-1}$. In particular, this combination procedure is performed five times for the dimer by adding potential barriers from left to right. %A part of the analytical expression of the reflection amplitude (equation~(\ref{eq2:reflection})) of the dimer is

\end{document}